\documentclass{aa}  

\usepackage{graphicx}
\usepackage{txfonts}
\begin{document} 

\def\muzero{$\mu_0$}
\def\msunyr{M$_{\odot}$ yr$^{-1}$}
\def\halpha{H$\alpha$}

   \title{The properties of the Malin 1 galaxy giant disk:}
\subtitle{A panchromatic view from the NGVS and GUViCS surveys}
   \author{S. Boissier\inst{1}, A. Boselli\inst{1}
          \and
L. Ferrarese\inst{2}, P. C\^ot\'e\inst{2}, Y. Roehlly\inst{1}, S.D.J. Gwyn\inst{2}
\and
J.-C. Cuillandre\inst{3}
\and
J. Roediger\inst{4}
\and
J. Koda\inst{5,6,7}          
\and
J.C. Mu\~nos Mateos\inst{8}          
\and
A. Gil de Paz\inst{9}
\and
B.F. Madore\inst{10}
}
   \institute{Aix Marseille Universit\'e, CNRS, LAM (Laboratoire d'Astrophysique de Marseille) UMR 7326, 13388, Marseille, France\\
              \email{firstname.lastname@lam.fr}
         \and
National Research Council of Canada, Herzberg Astronomy and Astrophysics Program, 5071 West Saanich Road,
Victoria, BC, V9E 2E7, Canada
\and 
CEA/IRFU/SAp, Laboratoire AIM Paris-Saclay, CNRS/INSU, Université Paris Diderot, Observatoire de Paris, PSL Research University, F-91191 Gif-sur-Yvette Cedex, France CEA/IRFU/SAp, Laboratoire AIM Paris-Saclay, CNRS/INSU, Université Paris Diderot, Observatoire de Paris, PSL Research University, F-91191 Gif-sur-Yvette Cedex, France 
\and
Department of Physics, Engineering Physics \& Astronomy, Queen's University, Kingston, Ontario, Canada
\and
NAOJ Chile Observatory, National Astronomical Observatory of Japan, Joaquin Montero 3000 Oficina 702, Vitacura, Santiago 763-0409, Chile
\and
Joint ALMA Office, Alonso de Cordova 3107, Vitacura, Santiago 763-0355, Chile
\and
Department of Physics and Astronomy, Stony Brook University, Stony Brook, NY 11794-3800, USA
\and
European Southern Observatory, Alonso de Cordova 3107, Vitacura, Casilla 19001, Santiago, Chile
\and
Departamento de Astrof\'isica, Universidad Complutense de Madrid, Madrid 28040, Spain
\and
Observatories of the Carnegie Institution for Science, 813 Santa Barbara Street, Pasadena, CA 91101, USA
             }

   \date{2016}

 
  \abstract
   {Low surface brightness galaxies (LSBGs) represent a significant percentage of
 local galaxies but their formation and evolution remain elusive. They may hold 
crucial information for our understanding of many key issues (i.e., census of baryonic and dark matter, 
star formation in the low density regime, mass function). The most massive examples --- the
so called giant LSBGs --- can be as massive as the Milky Way, but with this mass being distributed in a
much larger disk.}
   {Malin 1 is an iconic giant LSBG --- perhaps the largest disk galaxy known. We attempt to
bring new insights on its structure and evolution on the basis of new images covering a wide range in
wavelength.}
   {We have computed surface brightness profiles (and average surface brightnesses in 16 regions of interest), in six 
photometric bands ($FUV$, $NUV$, $u$, $g$, $i$, $z$). We compared these data to
various models, testing a variety of assumptions concerning the formation and evolution of Malin\,1.}
   {We find that the surface brightness and color profiles can be reproduced by
a long and quiet star-formation history due to the low surface density;
no significant event, such as a collision, is necessary.
Such quiet star formation across the giant disk is obtained in a
disk model calibrated for the Milky Way, but with 
an angular momentum approximately 20  times larger. Signs of small variations of the star-formation
history are indicated by the diversity of ages found when different regions within the galaxy are intercompared.}
   {For the first time, panchromatic images of Malin 1 are used to constrain the stellar populations and the history 
of this iconic example among giant LSBGs. Based on our model, the extreme disk of Malin 1 is found to have a
long history of relatively low star formation (about 2 \msunyr). Our model allows us to make 
predictions on  its stellar mass and metallicity.}


   \maketitle
%

\section{Introduction}

Low surface brightness galaxies (LSBGs)  are defined as galaxies 
with a disk central surface brightness (\muzero) much fainter than the typical  
\citet{freeman70} value for  disk galaxies  ($\mu_{0,B}$= 21.65 mag arcsec$^{-2}$)
with a limiting threshold variable depending on authors.
Galaxies with \muzero{} in the range 22 to 25 mag arcsec$^{-2}$ contribute 
up to 50\% of the light emitted by galaxies
according to \citet{impey97}. LSBGs thus represent a very significant percentage of 
local galaxies \citep[see also][]{oneil00} making them important 
contributors to the baryon and dark matter mass budget.
An incomplete census of this population could affect the shape of the luminosity function 
\citep{blanton05} and they should be considered when addressing the issues of the 
missing baryons problem,  and the population of quasars 
absorbers \citep[see][and references therein]{impey97}.
Despite their potentially important role, the origin and evolution of LSBGs 
have remained largely obscure, with their exceedingly low surface brightness hindering in-depth studies.

Different classes of galaxies can be found among LSBGs.  
While dwarf galaxies frequently satisfy the definition based on \muzero, 
a potentially distinct population of 
giant LSBGs also exists, having HI masses as high as $\sim 10^{10} M_{\odot}$ and 
rotation velocity upward of $\sim$ 200 km/s  \citep{pickering97,matthews01}. 
Such galaxies are extreme test cases that can help us solve fundamental issues 
concerning galaxy formation,  especially the angular momentum catastrophe and
the overcooling problem found in numerical models.
Various properties of this class of galaxies  have been studied in few papers targeting 
typically 10 to 20 galaxies, but many giant LSBGs may be missing from our galaxy catalogs \citep{impey89}.
Actually, in the nearby universe, new diffuse galaxies characterized by their 
low surface brightness are still being discovered 
owing to deep observations and improved detectors and techniques \citep[e.g.,][]{koda15,mihos15,vandokkum15,hagen16}.

Among the proposed scenarios for their formation and evolution, 
LSBGs  could result from peculiar initial conditions: 
disks with larger than average specific angular momentum have a longer scale-length for the
distribution of matter and as a result, a different star-formation history \citep{jimenez98,boissier03lsb}.
However, \citet{mapelli08} reminded us that it is hard for a disk with a 
large angular momentum
to survive in our cold dark matter dominated cosmology 
and its hierarchical formation history.
\citet{mapelli08} proposed instead that ring galaxies 
(such as the famous Carwheel galaxy) evolve into giant LSBGs.

Additionally, LSBGs allow the study of star formation in the low density regime, 
for which many issues are under active debate (i.e., lower efficiency, threshold, initial mass function -IMF- variations) as
demonstrated by the rich literature that followed the discovery of eXtended UV (XUV) galaxies with GALEX
\citep{gildepaz05,thilker07},
and the finding of  star formation where little was expected \citep{ferguson98,boissier07,goddard10,koda12}. 
Star formation in LSBGs has been studied \citep{boissier08,wyder09}
showing a small amount of dust and a low star-formation efficiency, as is generally found
in the outer low density regions in normal galaxies \citep{bigiel10} or in the low density gas deposited outside of 
galaxies by ram-pressure stripping \citep{boissier12,vollmer12,verdugo15}.

At a distance of 366 Mpc (from the NASA Extragalactic Database, based on its 
recessional velocity of 24750 km/s), 
Malin 1 is the prototype of giant LSBGs. 
It was first discussed  by 
\citet{bothun87} who showed that it has a very low surface brightness disk 
($\mu_{0,V}$= 25.7 mag arcsec$^{-2}$) and a prominent bulge.
Moore \& Parker (2006) obtained a deep $R$ band image, with detection up to 120 kpc from the center. 
Recent HST observations of the central kpc suggest that the part initially considered as a "bulge" 
is actually a normal SB0/a galaxy with a small bulge, a bar, and a high surface brightness disk with a spiral structure 
\citep{barth07}. 
The giant disk around it would then be an extreme case of the recently
discovered anti-truncated XUV disks \citep{gildepaz05,thilker05}.
Malin 1 clearly enters the class of giant LSBGs, with integrated quantities similar to usual
spirals, such as a large reservoir of neutral gas \citep[log($M_{HI}/M\odot$)= 10.6 - 10.8][]{matthews01,lelli10}, and 
a star-formation rate around 1 \msunyr{} based on its UV emission \citep{boissier08}.
Very recently, \citet{galaz15} published new $g$- and $r$-band images of Malin 1 obtained with 
the Magellan Clay telescope. 
These deep images allowed them to study the morphology of the extended disk  
for the first time. They clearly reveal the presence of a large scale system of
spiral arms, and some diffuse light, that they associate to possible past
interactions.
\citet{reshetnikov10} have indeed suggested that interactions of Malin 1 with a small companion, Malin 1B, now 
14 kpc away from its center, may be responsible for some of the morphological features of Malin 1. They also proposed 
that  another galaxy in the area, 350 kpc away from Malin 1 may have interacted with the LSBG 1 Gyr ago. 
Interactions and companions, however, are not limited to or typical of LSBG that live in less dense environments than
their high surface brightness counterparts \citep{rosenbaum04,rosenbaum09}.

Malin 1 happens to be projected behind the Virgo cluster, and was therefore 
imaged as part of the Next Generation Virgo Cluster Survey (NGVS, Ferrarese et al. 2012),
\nocite{ferrarese12} 
a large CFHT program providing very deep $u$,$g$,$i$,$z$ imaging. It was also imaged by the 
 GUViCS survey \citep{boselliguvics} providing GALEX $FUV$ and $NUV$ images.
This paper presents an analysis based on these six NGVS and GUViCS images. It was performed in parallel to 
the recent analysis of \citet{galaz15} who studied the spectacular morphology of the galaxy.
While we will compare our results to theirs when pertinent, our data cover the full spectral
energy distribution (SED) from the far ultraviolet to near-infrared, allowing us to also
study the stellar populations and their distribution in the extended disk of Malin 1.

In Section \ref{secobs}, we present the new data used in this paper, as well as some ancillary 
information collected for the study. The extraction of surface brightness profiles in six bands is discussed
in Section \ref{secsurfbrightprofile}. In Section \ref{secregions} we also extract the average surface brightness of
16 regions of interest to bring complementary information. The obtained SEDs are fit in simple
ways to get a rough idea of  the properties of the regions.
All the information obtained  is then used to discuss the formation and evolution of the giant disk of Malin 1 in Section \ref{secdiscussion}
where we test the predictions of several models.
We conclude in Section \ref{secconclu}.

%

\begin{figure}
\centering
\includegraphics[width=9.cm,clip]{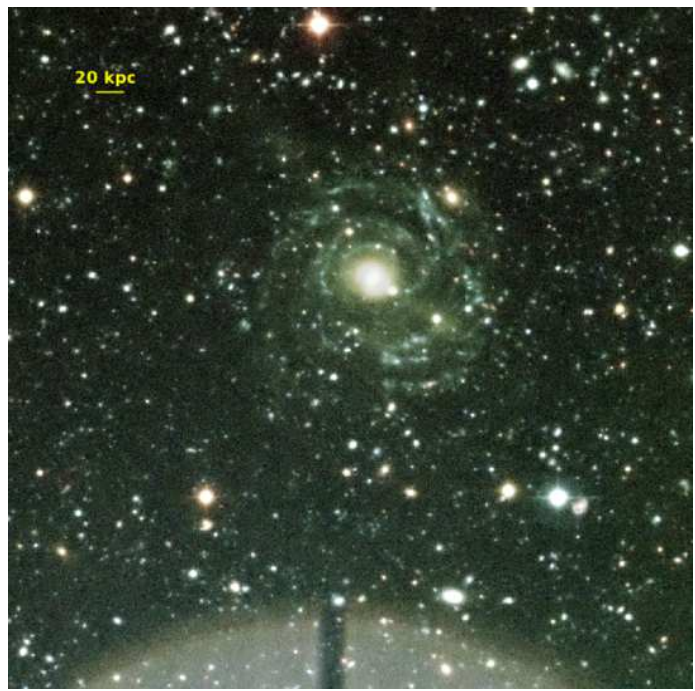}
\caption{Color composite of the NGVS Malin 1 images 
($u$, $g$, $i$, $z$ in blue, green, yellow, red, respectively).
The image measures about 4.5 arcmin across  (1 arcmin $\sim$ 107 kpc).}
\label{FigObsNGVSnative}
\end{figure}

\begin{table}
\caption{\label{tabexptime}Summary of the new observations presented in this work.}
\begin{tabular}{l l l l }
\hline
\hline
Filter & Wavelength  & Exposure  & PSF \\
        &  (\AA)           &   time (s)   &  (arcsec) \\
\hline
GALEX $FUV$  & 1528 &  5030           & 5.00 \\
GALEX $NUV$ &  2271 & 4662            & 5.00  \\
CFHT MegaCam $u$  &  3550 & 6402 &  0.79\\
CFHT MegaCam $g$  & 4750 & 3170   & 0.85\\
CFHT MegaCam $i$   & 7760 & 2055   & 0.54\\
CFHT MegaCam $z$  & 9250  & 3850  & 0.59\\
\hline
\end{tabular} 
\end{table}

\section{Observations}
\label{secobs}
\subsection{New UV and optical images}

In the $NUV$ and $FUV$ bands of GALEX, we use images from 
the  GALEX Ultraviolet Virgo Cluster Survey, GUViCS  \citep{boselliguvics}.
The typical exposure time of one GALEX orbit (1500s) corresponds to 
a surface brightness limit of about 28.5 mag arcsec$^{-2}$.
While \citet{boissier08} computed UV integrated magnitudes for 
Malin 1 using these data, in this paper we combine
all GALEX tiles in which the galaxy is imaged to obtain
an accumulated exposure time 3 times longer than in this previous work.
To do so, we used the Montage software \citep{jacob2010} 
and resampled the data on same grid and pixel scale as the MegaCam 
images \citep[in a similar manner as in][]{boissier15}, to facilitate the comparison between the UV and optical data. 
The GALEX PSF of the images is about 5 arcsec.  
Few foreground stars are present at these wavelengths, and they were manually masked.
The value of masked pixels was  estimated by interpolation from nearby pixels. 

We also use images obtained by the NGVS 
\citep{ferrarese12}, in the  $u$, $g$, $i$ $z$ bands. The survey reaches 29 mag arcsec$^{-2}$ in $g$.
A color composite of these bands is shown in Fig. \ref{FigObsNGVSnative}. 
Appendix \ref{secallimages} shows all the individual images used 
in this work at their native resolution.
The NGVS images were processed with the Elixir-LSBG pipeline 
optimized for the recovery of low surface brightness features 
(see Ferrarese et al. 2012 for details).
A bright foreground star is present at the South of the galaxy,
 producing a large halo (partly visible in Fig. \ref{FigObsNGVSnative}) and
causing a large-scale low surface brightness structure in the background. 
We fit a  radial gradient to this structure, and subtracted it 
to flatten the images before analysis.
\label{secremovestar}
\label{secreduc}
We then used the point source catalog maintained by the NGVS collaboration 
to prepare a mask from the positions of objects marked as stars. 
This mask was visually inspected within the perimeter of the galaxy to mask 
additional stars and background galaxies or to unmask  HII regions 
from the galaxy. In case of doubt, the objects were not masked.
Masked areas were then replaced by interpolated values from the surrounding pixels.

Table \ref{tabexptime} lists the final exposure times of the images, with their spatial resolution
(about 5 arcsecs in the GALEX images, and measured on some of the numerous stars surrounding the galaxy
in the NGVS images).

The NGVS images were  convolved with a Gaussian filter in order 
to match the spatial resolution of the GALEX data, ~ 5 arcsecs
(all the results published thereafter are obtained within bins or regions larger than this size). 
Surface photometry (radial profiles and regions) was measured in these images, after 
correcting for  a foreground Galaxy extinction with $A_V$=0.101 mag
\citep[from][ as found on the NASA Extragalactic Database]{schlafly11}.
The sky level and dispersion was measured in all cases in
a large number of small regions surrounding the galaxy.

\begin{figure*}
\centering
\includegraphics[width=18.cm,clip]{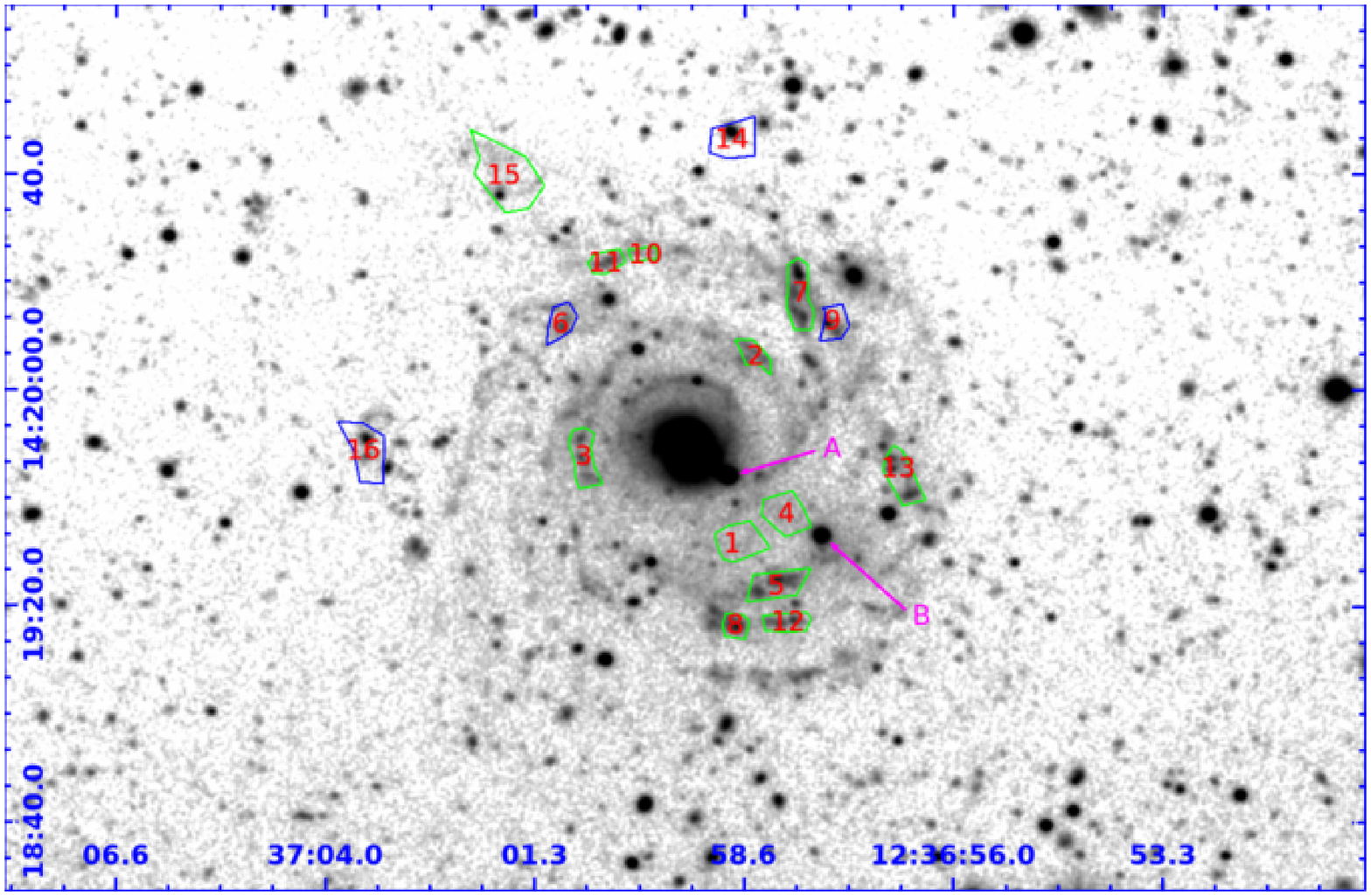}
\caption{$g-$band image of Malin 1. 
Labels A and B, respectively, mark the position of Malin 1B, the nearby companion galaxy of 
Malin 1 identified by \citet{reshetnikov10} and of a likely background galaxy \citep{galaz15}.
A few regions selected for additional analysis (see Sect. \ref{secregiondef}) are 
also shown (green delimited regions were selected in the $g-$band images, 
blue delimited regions were chosen in addition from the UV images, after 
the reduction described in Sect. \ref{secreduc}).
All the original images at their native resolution are shown in appendix \ref{secallimages}.
\label{FigObsNGVS}}
\end{figure*}

\label{secimages}
A simple examination of the final images already provides us with information concerning Malin1.
First, the extended disk presents a clear spiral structure in its bluer bands. It is apparent in 
the $g$ band (Fig. \ref{FigObsNGVS}), confirming the very recent result of \citet{galaz15}. 
The spiral arms are similarly striking in the $u$ band.  As \citet{galaz15}, we also observe several diffuse areas
such as the region marked 15 in Fig. \ref{FigObsNGVS}.
%
Region 1 has a brighter emission than the other inter-arm areas. It thus
could be a feature produced during an interaction with Malin 1B that is not very far. 
Region 4, just next to it presents a similar level
of emission. It is, however, located next to a background galaxy. 
\citet{galaz15} suggest that the elongated structure in the direction of region 4 is actually part of the 
background galaxy. Notice that they used specific data reduction techniques to enhance the detection 
of faint structures, and we refer to their work for a detailed morphological analysis. 

Extended emission is also clearly seen in the $FUV$ and $NUV$ images (appendix A) even if 
the spiral arms are not as visible as in the optical, partly because the data are not as deep, and have a lower spatial resolution. 
Some areas emitting in the $u$ or $g$ band are clearly not emitting in $FUV$,
and are weak (at best) in NUV. This could be a sign of episodic star formation. 
Regions formed less that 100 Myr still have massive stars emitting in $FUV$, while 
older regions no longer contain UV-bright stars.

\subsection{Ancillary data concerning the gas and dust content of Malin 1}

\citet{lelli10} have re-analyzed the \citet{pickering97} VLA observations of Malin1 to 
produce a HI surface density profile that we adopt for comparison to our models. We simply apply a 1.4 
correction factor
to convert it to the total neutral gas surface density. 
They also provide a rotation curve (albeit subject to uncertainties due to the inclination of the galaxy as will be discussed in the next section).
Malin 1 was never detected in molecular gas tracers \citep{braine00,lee14}. We thus assume the molecular fraction
is low in this very extended low density disk, as may be expected from the extrapolation
of the trends found in nearby galaxies \citep{leroy08} or in other low density regions \citep{dessauges14}. 
For these reasons, we do not correct the gas profile for inclusion of molecular gas,  
bearing in mind this introduces a large uncertainty.

Proper estimations of the amount of dust attenuation or of the dust mass require 
the measurement of the  far-infrared total emission  \citep[e.g.,][]{cortese08}.
\citet{rahman07} attempted to observe Malin 1 with \emph{Spitzer} and obtained only upper 
limits at the longest wavelengths. They did have a detection at 24 microns, but their
image shows that it only concerns the central part of the galaxy. 
The Virgo cluster was observed with \emph{Herschel} in the context of the HeViCS survey \citep{davies10}. 
We inspected the images of this survey. Malin 1 is close to the edge of the survey but is within the observed area at 250 and 500 microns. Nothing is clearly detected in either of the corresponding images. 
The limits from the \emph{Spitzer} observations  were combined with 
integrated UV luminosity by \citet{boissier08} who estimated the FUV attenuation to be lower than 0.4 mag. LSBGs in general
are believed to contain less (or colder) dust than their high surface brightness counterparts \citep{rahman07,hinz07,das06}

%

\section{Surface brightness profiles}
\label{secsurfbrightprofile}

\subsection{Geometrical parameters}
\label{secgeompar}
\begin{figure*}
\centering
\begin{tabular}{ c c c  }
input image & model & residual \\
\includegraphics[width=4cm,clip]{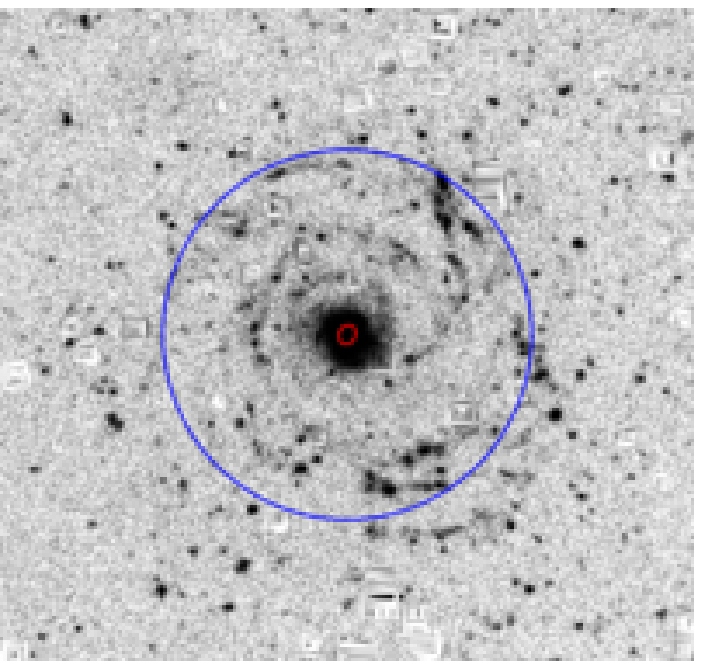}  & \includegraphics[width=4cm,clip]{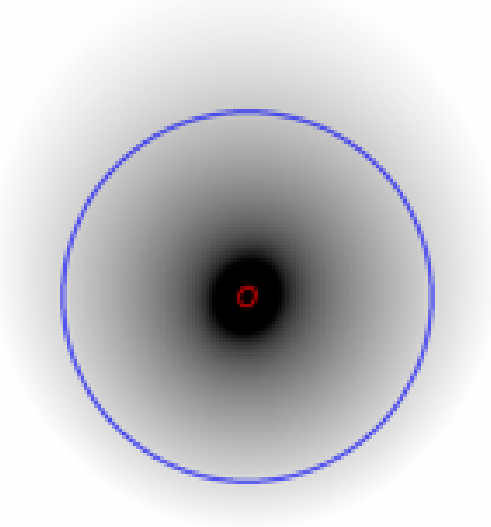}  & \includegraphics[width=4cm,clip]{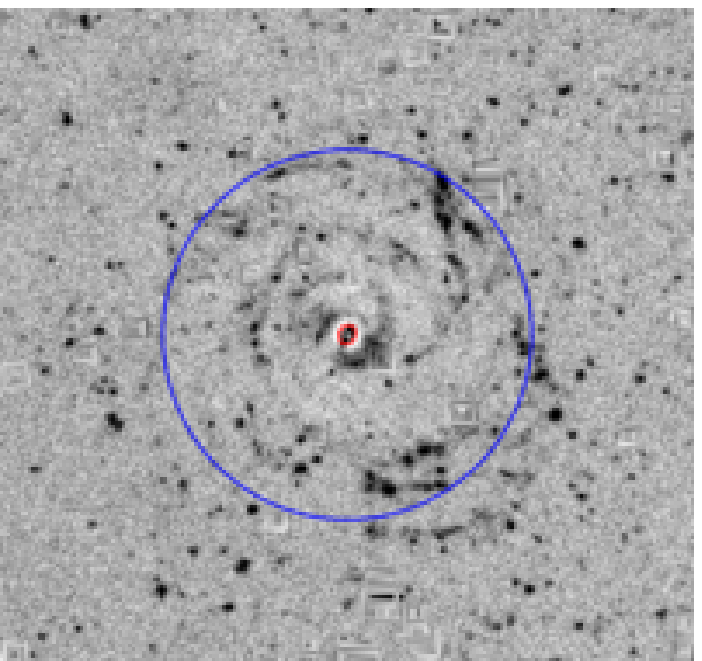}   \\ 
\includegraphics[width=4cm,clip]{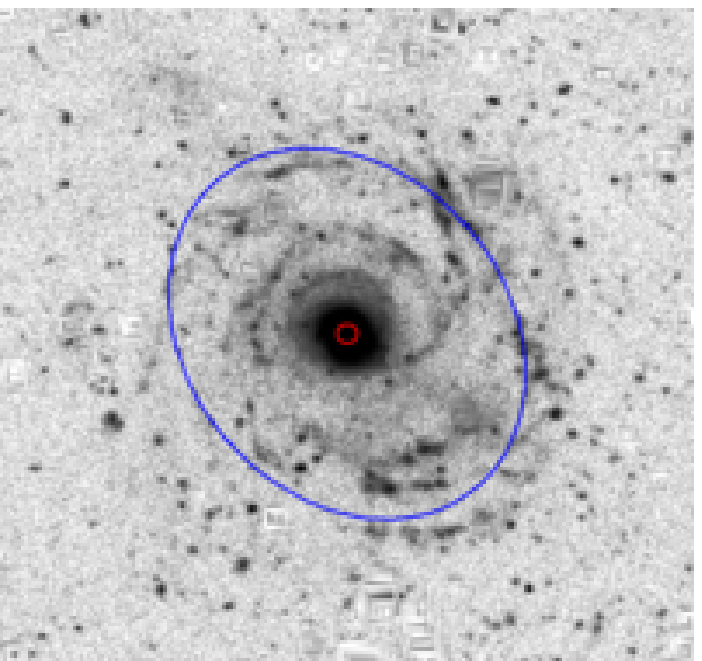}  & \includegraphics[width=4cm,clip]{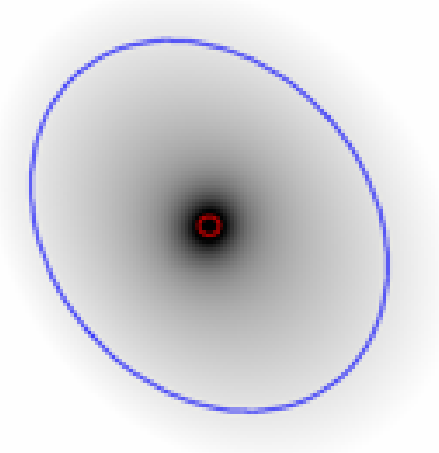}  & \includegraphics[width=4cm,clip]{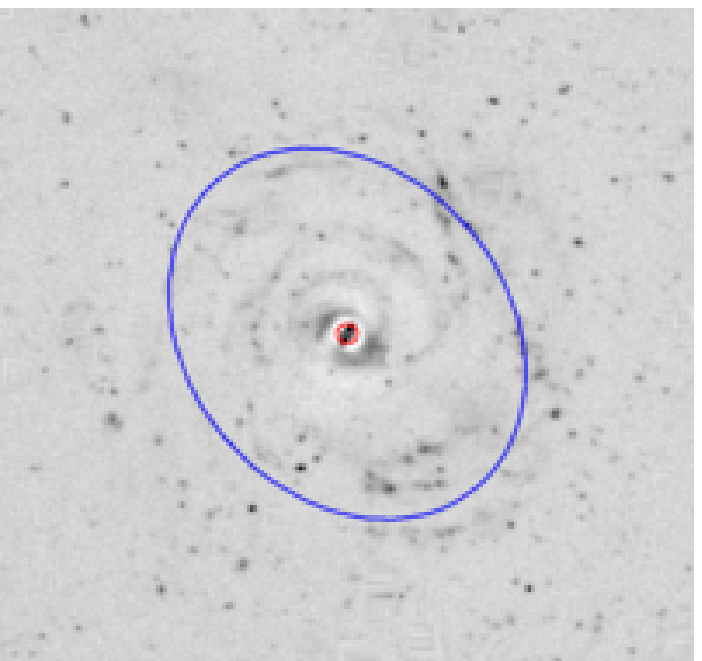}   \\ 
\includegraphics[width=4cm,clip]{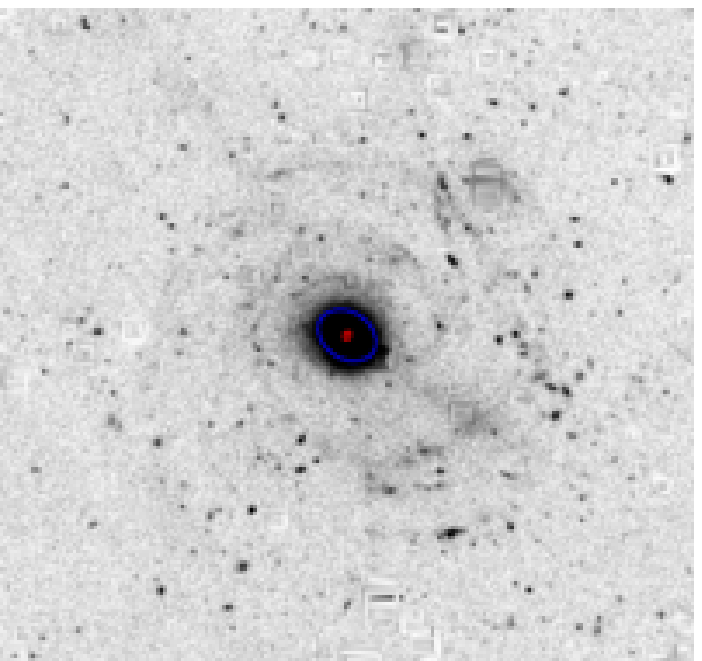}   & \includegraphics[width=4cm,clip]{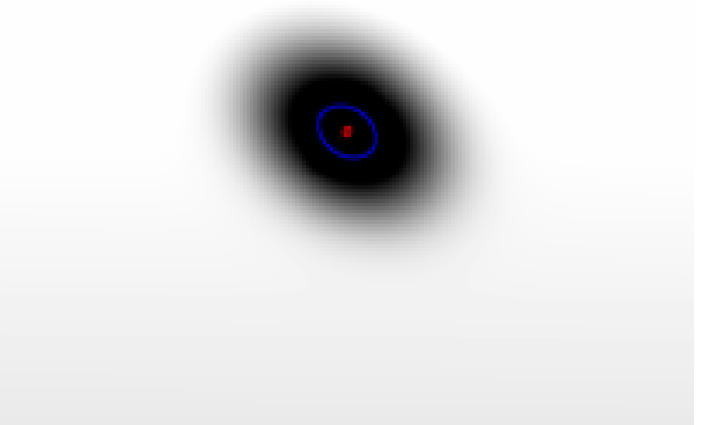}  & \includegraphics[width=4cm,clip]{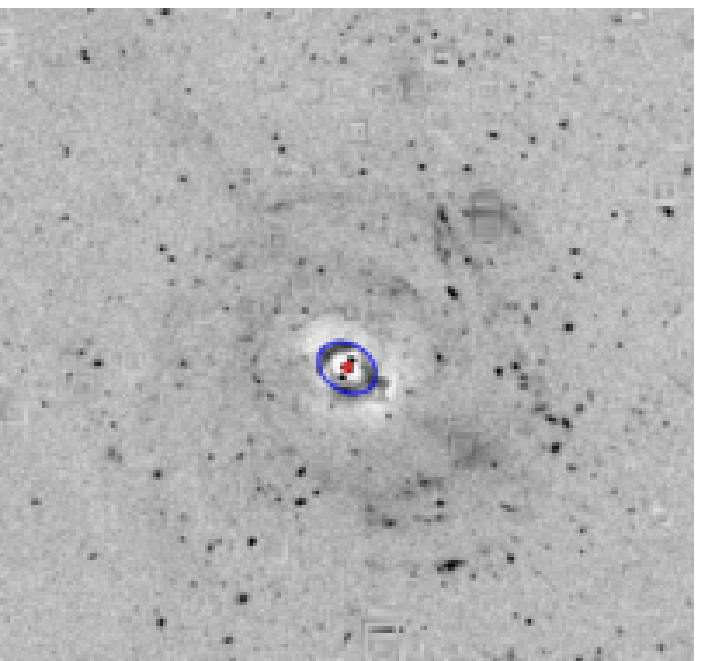}   \\ 
\includegraphics[width=4cm,clip]{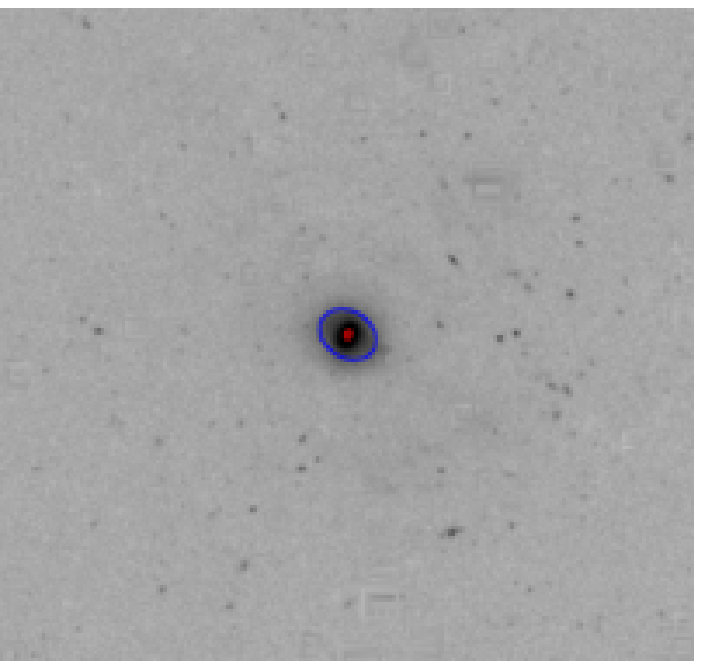}  & \includegraphics[width=4cm,clip]{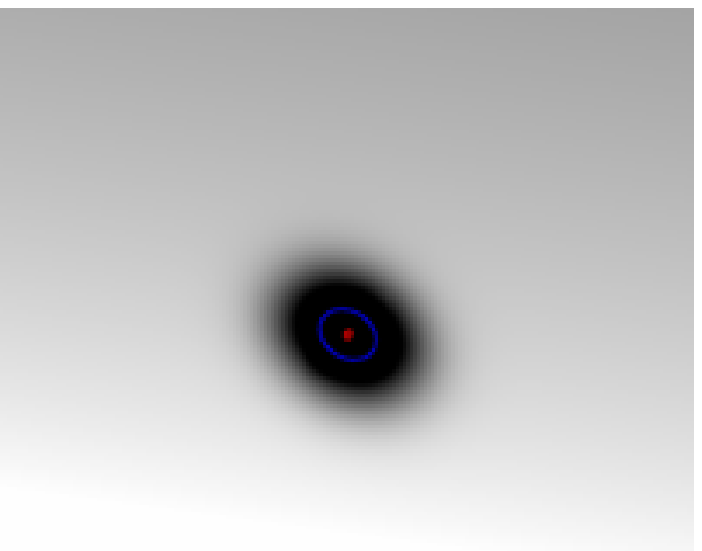}  & \includegraphics[width=4cm,clip]{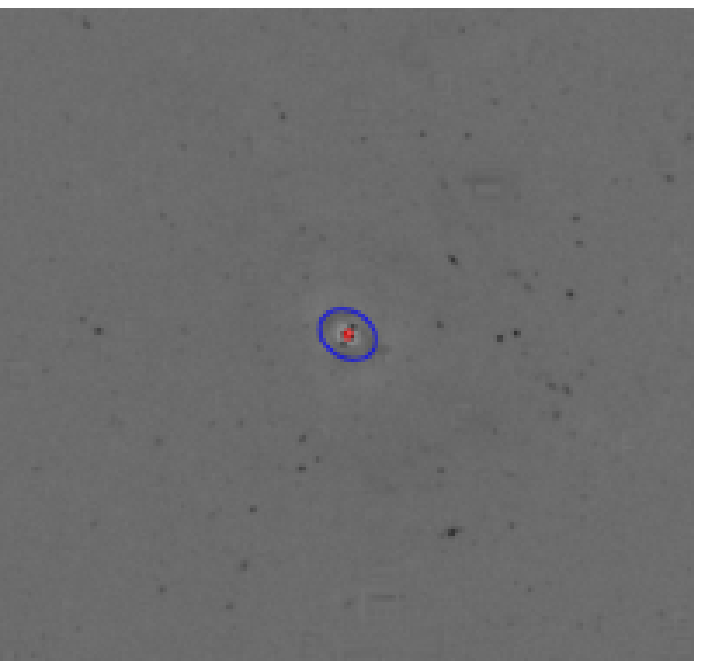}   \\ 
\end{tabular}
\caption{GALFIT results. On the left, we show the input image for GALFIT, in the middle its model, and on the right the residual. In each image, we show 2 ellipses indicating the geometry found by GALFIT in terms of effective radius, PA, axis ratio for the disk (outer blue ellipse) and the sersic component (inner red ellipse). From top to bottom: u, g, i z.
\label{FigGALFIT}
}
\end{figure*}

The surface brightness profiles can be measured easily once the geometrical parameters 
(inclination and position angle) are fixed. For Malin 1, values used as reference for these parameters
in the literature are from \citet{moore06} who matched manually an ellipse to their $R$ band 
image and found an axis ratio of 0.8 and a
position angle of 43 $\deg$.
It is possible to revisit the values of these parameters with our deep NGVS images. To do so, we used GALFIT
\citep{peng02} to fit 3 components
(Sersic, exponential disk and a flat sky gradient) to each of the optical images. The  center of the Sersic and exponential disk
components were fixed and coincide with the peak of emission in the center of the optical images. 
Initial values were visually guessed for other parameters (position angle, axis ratio, effective radius) but varying them slightly does not change the results. 
GALFIT independently fits the three components in each band. 
The geometrical results of GALFIT are given in Table \ref{tabgeom} with the uncertainties, as provided by GALFIT. We omit the sky component that is almost null and close to flat, but that we included in the fit in order to take into account possible residuals from the procedure we applied to subtract the background (especially the contribution from a nearby bright star,
see Sect. \ref{secremovestar}). 
The resulting fits can also be seen in Fig. \ref{FigGALFIT}

The results indicate that our simple fit finds an extended disk in the bluest bands ($u$, $g$), but fails to do so 
in the reddest bands ($i$, $z$) in which the extended disk (and the associated spiral arms) are less  conspicuous and GALFIT instead fits inner structures that are more prominent.
It is also clear from the residuals that the Sersic component fails to fit the inner region.
This is not surprising considering the different structures found by \citet{barth07}: the bulge, bar, and disk that we have not attempted to reproduce. This may partly affect the parameters found for the extended disk even if it is well seen in the model images in the $u$ and $g$ bands.
Technically, it would be possible to force GALFIT to fit more components (e.g., inner bulge, bar, and disk, spiral structure), or to keep a large radius for the disk component, but the rest of the paper 
does not depend strongly on these results 
(except for deciding on an axis ratio and PA for the ellipse procedure as detailed later). 
The outer disk is very weak in $z$ band and a proper determination would be in any case difficult. 
The PA and axis ratio found in the $u$ and $g$ bands are very similar to those of \citet{moore06}, even if the fit is consistent with a face-on orientation in $u$.
The difference in apparent inclination could be related to 
the respective weight of the geometry of the spiral arms of the 
star forming extended disk, and the distribution of the old stellar 
populations, although it is probably affected by the different  
signal to noise in the two bands.
The formal errors found by galfit are small, but they do not necessarily imply this solution is definitive. As a test,
we attempted to fit respectively the $u$ and $z$ bands with the axis ratio and position angle
derived from respectively the $z$ and $u$ 
bands. We found that GALFIT finds a solution in both cases without a significant increase of the $\chi^2$:  
the range of values found for the PA and axis ratio at various wavelengths are thus all consistent with our data.
As a result, the real inclination is somewhat unconstrained. In this work, we computed surface 
brightness values for two inclination, one matching the
geometry found in the $u$ band 
(showing well the extended star forming disk and the young stellar populations),
and one corresponding to the geometry in the $g$ band (older stellar 
population, optimal quality data).
\citet{lelli10} have found a warped HI disk in Malin 1. The very low surface density of the stellar disk, and the asymmetry of  the spiral arm (the most visible regions at large radii) would make difficult the detection of a stellar warp with our data. Nevertheless, we adopt a third possible assumption for the geometry taking into account this possibility, inspired by \citet{lelli10}, that is an axis ratio of 0.8 as previously, but a variable position angle (-1 degree for the inner 15 arcsecs, and rising with radius 
as 1.4 $\times$ the radius in arcsec beyond 15 arcsec, as given in Lelli et al. 2010).

\begin{table*}
\caption{\label{tabgeom}Galfit results.}
\begin{tabular}{l l l l l l l l l }
\hline
\hline
Filter & Disk                       &  Disk              &  Disk                          &          Sersic             &     Sersic              &       Sersic                     &   Sersic   & reduced  \\   
         & scalelength                & b/a                &   PA                         &    index            & effective radius            &  b/a              &  PA               & $\chi^2$ \\
                  & (arcsec)          &                      &  (deg)                                 &                         &  (arcsec)           &                     &  (deg)            &       \\
u               &  23.23 $\pm$  0.47 &  0.97  $\pm$ 0.01 &   N\/A                       &   2.71 $\pm$ 0.03   & 2.07 $\pm$ 0.02   & 0.85 $\pm$ 0.01   & -24.00 $\pm$ 1.14 & 1.148  \\    
g               &  25.26 $\pm$ 0.26  &  0.80  $\pm$ 0.01 &   40.26 $\pm$ 0.82 &   2.95 $\pm$ 0.01   & 2.192 $\pm$ 0.01   & 0.93 $\pm$ 0.01   & -19.58 $\pm$ 0.99 & 1.155 \\  
i                &   3.88 $\pm$ 0.02   &  0.75  $\pm$ 0.01 &   57.07 $\pm$ 0.32 &   2.45 $\pm$ 0.01   &  0.96 $\pm$ 0.01 & 0.67 $\pm$ 0.01   & -25.42  $\pm$ 0.17 & 1.377 \\
z               &   3.74 $\pm$ 0.02   &  0.78  $\pm$ 0.01 &   56.79 $\pm$ 0.41 &   2.10 $\pm$ 0.01   &  0.95 $\pm$ 0.01    & 0.70 $\pm$ 0.01   & -23.86 $\pm$ 0.17 & 1.343 \\
\hline
\end{tabular} 
\end{table*}

\subsection{Measurement of the surface brightness  profiles}

\begin{figure*}
\centering
\includegraphics[angle=-90,width=18.cm,clip]{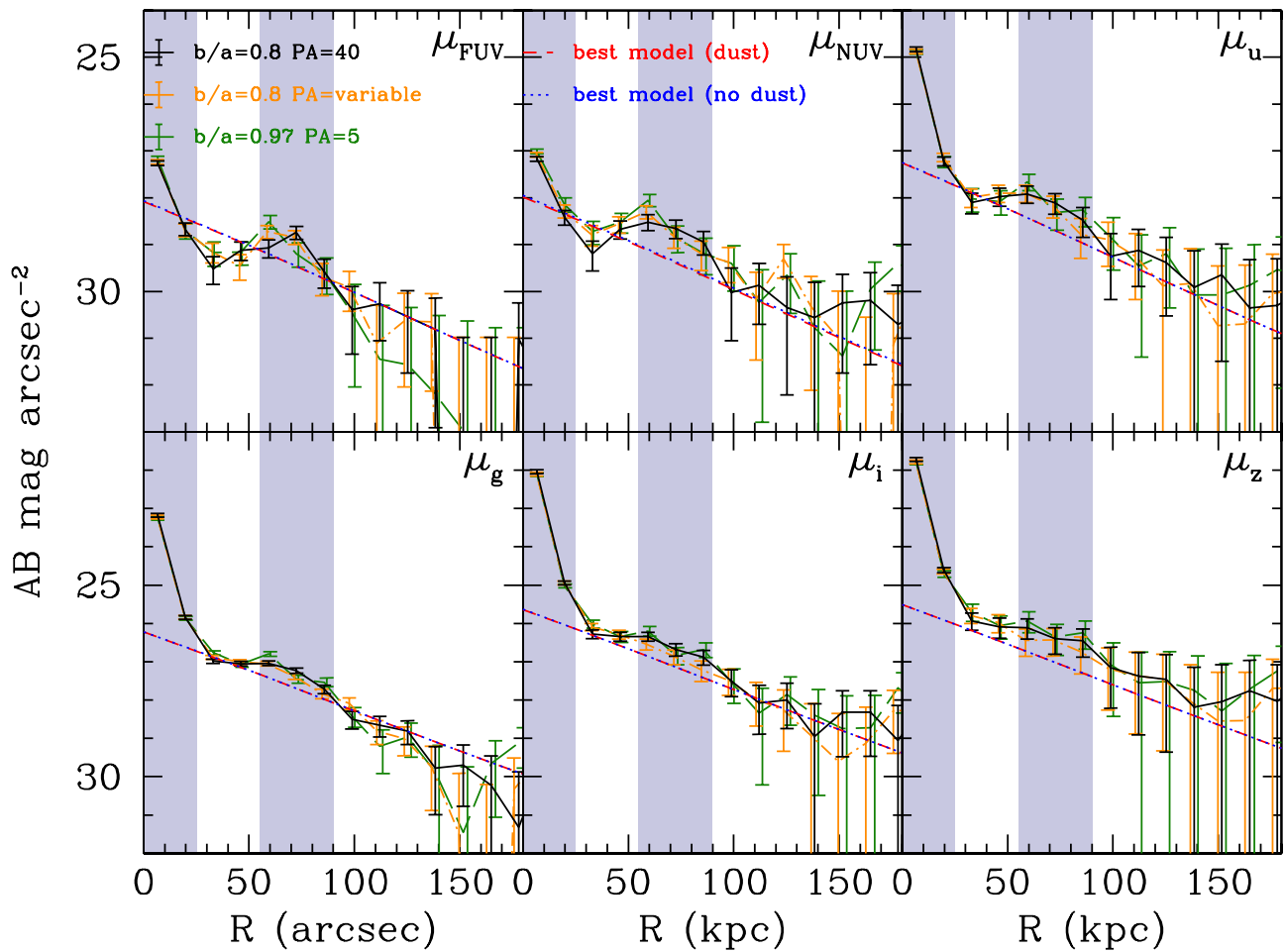}
\caption{Surface brightness profiles for Malin 1 assuming 
three possible geometries (see legend in top-left panel): 
an axis ratio of 0.8 and a PA of 40 degrees, an axis ratio of 0.97 and a PA of 5 deg, or an axis ratio of 0.8 and a variable PA as suggested in Lelli et al. (2010).
Profiles have been de-projected to their face-on value. They are compared to the best fit model of Sect. \ref{sublargespin} including or not dust attenuation.
The shaded area presents the regions excluded from the fit (avoiding the bulge and bumps likely related to the spiral structure) but the model is in agreement with the data also in the remaining of the disk (within a few sigma).
}
\label{FigObsProfiles}
\end{figure*}

Figure \ref{FigObsProfiles} shows the $FUV$ to $z$ radial profiles of Malin 1, assuming 
the three sets of assumptions concerning  the parameters (constant PA and slightly inclined with $b/a$=0.8,  almost face-on with $b/a$=0.97, and variable PA with $b/a$=0.8).
They are corrected for the $cos(inclination)$ factor to obtain face-on values. 
They were computed with the task ELLIPSE in IRAF, 
keeping the center fixed and computing the profiles in the 
same rings in each band. 
Error-bars were computed taking into account the uncertainty in the sky determination, combining
the  sky deviation itself (on a pixel basis), but also the deviation between several area of sky   
around the galaxy to account for residual structure in the 
background of the images, as presented in  \citet{gildepazmadore05}, which often
dominates the error budget at low surface brightness in our images.
The profiles present small differences but are mostly within the error-bars of each other. 
The real geometry may be a bit different from the 3 test-cases that we adopted, but 
our discussions and conclusions will remain unchanged in that case.
In our profiles in the $g$ and $i$ bands, we measure points above the noise level and with an approximate exponential 
radial distribution up to about 130 kpc.

\section{Spectral energy distribution of regions of interest}
\label{secregions}
\label{secSEDregions}

A galaxy SED (even when azimuthally averaged at a given radius) will be a mixture of multiple generation of stars. Ideally, the detailed history 
of individual pixels can be assessed in nearby galaxies. However, due to the very low surface brightness of most of the disk, the
signal per pixel is very low in a large part of the disk, making an analysis of the full map of the galaxy on a pixel basis difficult. 
In this section,  we define  16 distinct regions that are readily visible in the optical or UV images and encompass enough pixels to increase the
overall signal to noise ratio.
%
This approach allows us to establish that the regions have not been formed simultaneously; rather, several distinct stellar populations appear to be present in 
the low surface brightness disk of Malin 1.

\subsection{Measurements in 16 regions}
\label{secdefreg}
\label{secregiondef}

\begin{figure*}
\centering
\includegraphics[angle=-90,width=18.cm,clip]{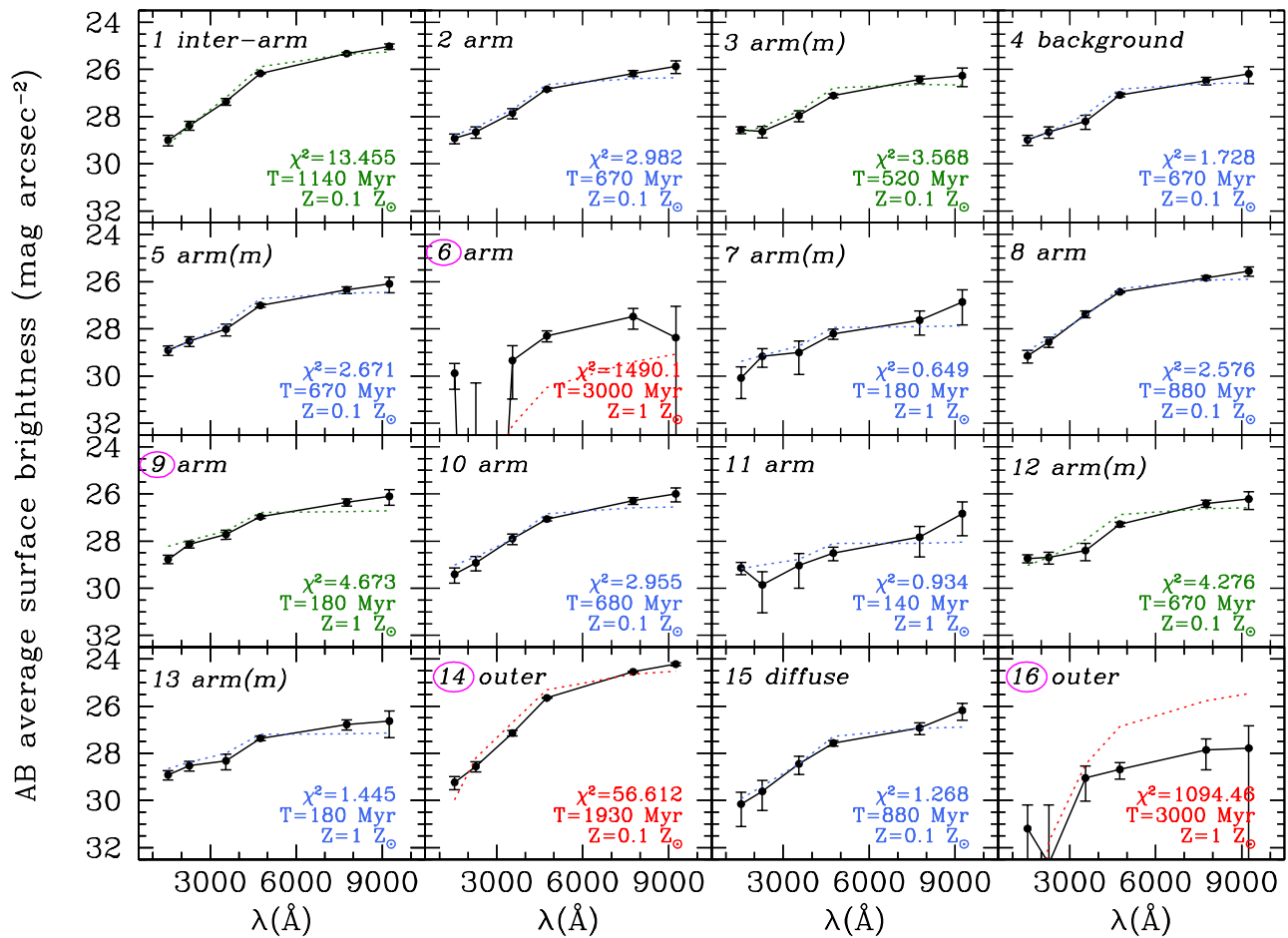}
\caption{$FUV$ to $z$ SEDs for 16 selected regions (defined in Fig. \ref{FigObsNGVS} and in Section \ref{secdefreg}). The black
lines with points and errorbars are the observed SEDs. The dotted line is the SED for the single generation fit (Sect. \ref{secsimplesed}).
The color indicates the quality of this fit: relatively good, average, bad fits are shown in blue, green and red, respectively, based on the  $\chi^2$ value given in each panel. 
The number id of each region is circled for regions selected in ultra-violet, the other regions were chosen in the optical. 
}
\label{FigSedRegions}
\end{figure*}

We defined 16 small regions manually by carefully inspecting the images at our disposal.
We have not attempted to catalog every single region detected in a particular image, but
have tried to cover the observed diversity of regions (spiral arms, inter-arm, diffuse area) and range in distance from the
galactic center. The shape and position of the regions are shown in Fig. \ref{FigObsNGVS}. 
Most regions were selected in the $g$ MegaCam image, many of them being
part of the spiral structure visible in this image and in the $u-$band 
(2, 3, 5, 11, 8, 10, 7, 12, 13). 
Some of the regions  (3, 5, 7, 12, 13)
encompass several knots, but the knots have similar color and appearance within the selected region in that case.
A few regions were selected as being diffuse in these images (1,4,15, 
some of them
possibly resulting from past interactions, or 
linked to a background galaxy as discussed in Section \ref{secimages}).
Some additional regions (6, 9, 14, 16) were chosen from the GALEX images on
the basis of their relative high $FUV$ emission, but they 
were not obvious choices in the optical bands and
had not been selected on their basis.
All regions were defined manually, using the images matched to a common resolution and with contaminants having been removed.
A  summary of the characteristics of the regions is given in Table \ref{tabregions}.

Aperture photometry was performed  directly on the NGVS and GUViCS images
(with masked objects interpolated over, and convolved to match the GUVICS resolution as for the surface brightness profiles) and we 
computed the average surface brightness in the 16 regions. The obtained SED are shown in  
Fig. \ref{FigSedRegions}.

\subsection{SED fitting}
\label{secsimplesed}

In an attempt to establish the age of the 
last significant star-formation event, we use the colors of an aging stellar population from \citet{boissier08}. They showed that the $FUV-NUV$ color
reddens quickly after a brief ($10^8$ yr) epoch of star formation stops, and that the colors of various LSBGs are consistent with
a succession of quiescent and active phases of star formation (the results were similar for a shorter or longer star forming episode). 
The same model predicts the magnitude in all bands, 
and instead of using only the $FUV-NUV$ color, we use here the full SED to try to constrain the age of the dominant population
of each region. As in \citet{boissier08}, we assume
a \citet{kroupa01} IMF, and that the last
significant star forming event lasted $10^8$ yr.
We interpolate metallicities in the range 
0.1 to 1 solar, and ages up to 3 Gyr. Older ages are not physical for this single 
population assumption
for which we assume the light is dominated by the "last" star forming event, 
neglecting the underlying 
population that may include a broader range of ages.

The best fits are shown in Fig. \ref{FigSedRegions} together with the $\chi^2$ values. Table \ref{tabregions} provides the results 
and the interval of confidence (one sigma). 
Globally, the metallicities are very poorly defined (the fit usually falls on one extrema, and all of the regions
but two are consistent with 0.1 solar). Metallicities derived from fitting  
broad-band observations are known to have limitations \citep[e.g.,][]{gildepaz02}.
Ages are better constrained
and are typically found in the range of a few hundred Myrs, with a clear spread (not all the regions in the spiral arms were
formed simultaneously). The regions selected from their $FUV$ emission are usually poorly fit.  
Inspection of the SED suggests a blue $FUV-NUV$ color but red SED 
otherwise that cannot be fit with a single stellar population, suggestive of multiple populations.

The diffuse region 15 is consistent with stars born 0.9 Gyr ago (between 0.5 and 1.3 Gyr). 
This is consistent with the age of the interaction proposed by \citet{reshetnikov10} but among the 
spread of ages found in the other regions.
The inter-arm regions 4 indicates a relatively young age (0.67 Gyr), while the nearby region 1 is older (1.14 Gyr), possibly indicating a different nature
\citep[region 4 could be part of a background galaxy according to ][]{galaz15}.

\begin{table*}
\caption{\label{tabregions}List of regions (ranked by distance to the center of the galaxy) and results of single age population fit.}
\begin{tabular}{l l l l l p{9cm}}
\hline
\hline
ID & $\chi^2$ & Age   & Metallicity & Selected in &Comment \\ 
    &  reduced  & (Gyr) & Solar        & (band) & \\
\hline
   1 &  13.455   &  1.14  ( 0.87 -   1.16 ) &  0.10  (   0.10 -   0.22 )       & $g$  &Diffuse inter-arm region \\ 
   2 &   2.982    & 0.67  (   0.58 -   0.82 ) &  0.10  (   0.10 -   0.23 )      & $g$  &Single knot, spiral arm \\ 
   3 &  3.568    & 0.52  (  0.17 -   0.68 ) &  0.10  (   0.10 -   1.00 )        & $g$  &Multiple knots, spiral arm\\ 
   4 &   1.728    & 0.67  (   0.46 -   0.77 ) &  0.10  (   0.10 -   0.36 )     & $g$  &Diffuse inter-arm region (background?) \\ 
   5 &   2.671    & 0.67  (   0.48 -   0.75 ) &  0.10  (   0.10 -   0.32 )      & $g$  &Multiple knots, spiral arm\\ 
   6 & 1490  &   3.00  (   2.72 -   3.00 ) &  1.00  (   0.99 -   1.00 )        & $FUV$&Single knot, spiral arm\\ 
   7  &  0.649    & 0.18  (   0.12  -  0.95 ) &  1.00  (   0.10 -   1.00 )      & $g$  &Multiple knots, spiral arm\\ 
   8 &   2.576    & 0.88   (  0.78 -   0.93 ) &  0.10  (   0.10-   0.20 )       & $g$  &Single knot, spiral arm\\ 
   9  &  4.673    &  0.18  (   0.17-   0.68 )     &  1.00  (   0.10 -   1.00 ) & $FUV$ &Single knot, spiral arm\\ 
  10  &  2.955    & 0.68  (  0.57 -   0.95 ) &  0.10  (   0.10 -   0.34 )       & $g$  &Single knot, spiral arm \\ 
  11  & 0.934    & 0.14  (  0.06 -   0.41 ) &  1.00  (   0.10 -   1.00 )       & $g$  &Single knot, spiral arm\\ 
  12 &   4.276    & 0.67  (   0.41 -  0.73 ) &  0.10  (   0.10 -   0.33 )       & $g$  &Multiple knots, spiral arm\\ 
  13  &  1.445    & 0.18  (   0.15  -  0.65 ) &  1.00  (   0.10 -   1.00 )     & $g$  &Multiple knots, spiral arm\\ 
  14 &  56.612    & 1.93  (  1.84 -   1.94 )     &  0.10  (  0.10 -   0.11 )  & $FUV$&Single knot, outer region\\ 
  15 &   1.268    & 0.88   (  0.51 -   1.31 ) &  0.10  (   0.10  -   0.61 )     & $g$ &Diffuse region \\ 
  16 & 1094  &   3.00  (   2.92 -   3.00 ) &  1.00  (   1.00 -   1.00 )        & $FUV$&Single knot, outer region\\ 
\hline
\end{tabular}
\tablefoot{The intervals indicated between parenthesis are computed on the basis of the $\chi^2$  following Avni (1976). 
They provide the 68\% confidence interval (unless the limit of the range investigated is reached, in this case, the limiting value is the limit of the explored range since all the explored values are within this 68\% confidence).}
\end{table*}
\nocite{avni76}

\label{secburst}
The approach based on this single stellar population is probably not realistic but is very simple
(only 2 parameters). 
Although full star-formation histories cannot be derived using only six bands, we attempt in 
the appendix \ref{secgazparfit} to fit the SEDs with  (slightly) more complex star-formation histories, using
SED fitting tools available online through the GAZPAR interface\footnote{\url{http://gazpar.lam.fr/}}. 
We also found with this method a diversity of ages or star-formation histories among the regions.

To summarize this section, the SED and the ages that we have derived using different approaches  
do not indicate a unique behavior. The age is not peaked at an unique value as would be the case if 
all (most) of the regions formed simultaneously, as would be if they were the result of a short-lived 
interaction between the galaxy and a companion.
Of course, this concerns the behavior on relatively short-time scale of individual regions, and we will investigate in Sect. \ref{sublargespin} the long-term smooth star-formation history on the basis of the radial profiles.


\section{Discussion}

\begin{figure}
\centering
\includegraphics[width=9.cm,clip]{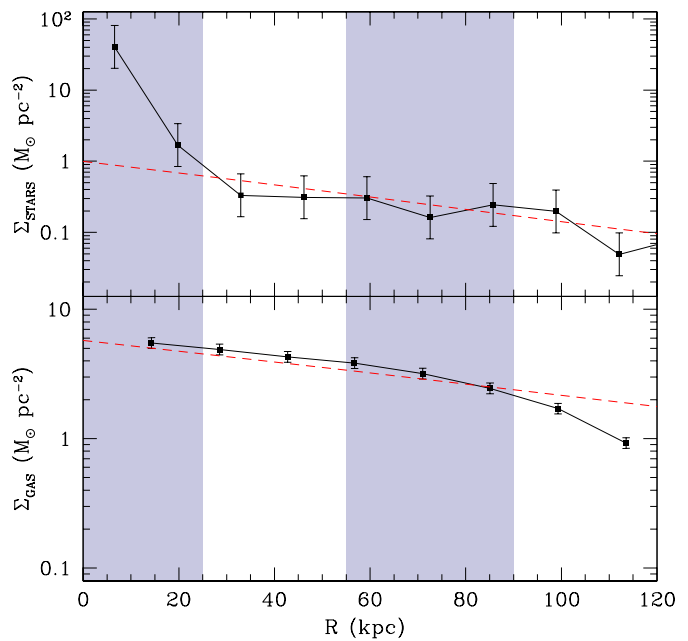}
\caption{Top: Stellar mass density profile (points) computed on the basis of the surface brightness profiles (see Sect. \ref{joel}).
The error-bars indicate a factor two. Many uncertainties affect this profile such as the photometric errors, but also the
model-dependent fit assuming a family of star-formation history, or an initial stellar mass function, so the factor of two is purely illustrative.
Bottom: gas column density profile (points), converted from HI by multiplying by a factor 1.4 from \citet{lelli10}.
Error-bars are not provided by \citet{lelli10} so a
10 \% uncertainty has been assumed.
The shaded area indicates the regions excluded for the fit (based on the bulge and bumps observed in the photometric profiles) but the model is in agreement with the data also in the remaining of the 
disk. In both panels, the red dashed lines indicates our best model profile.
}
\label{FigGas}
\end{figure}

\label{secdiscussion}
\subsection{The large angular momentum hypothesis}
\label{sublargespin}

\subsubsection{Disk galaxy models}

It was proposed that LSBGs are analogous to high surface brightness
galaxies, but are simply more extended due to their larger angular 
momentum \citep{jimenez98}.
This assumption was tested, for instance, by
\citet{boissier03lsb} who modelled LSBGs with the same assumptions
as adopted for the  Milky Way and disk galaxies \citep{boissier99, boissier00},
except for their larger spin parameter (a measure of the specific angular momentum).
\citet{mcgaugh98} considered this possibility but found it unlikely to
match the surface brightness distribution and the fact that LSBGs reside in
low density environment.
\citet{boissier03lsb} nevertheless showed that several LSBG properties are in agreement with this
simple assumption.
Here, we adopt this family of models, in the version presented in \citet{munoz11},
with a \citet{kroupa01} IMF. The models depend only on two parameters,
the circular velocity ($V$) and the spin parameter ($\lambda$). They are characterized
by scaling relationships such that the total mass of a galaxy scales with $V^3$, and the scale-length 
with $V \times \lambda$. The evolution at several radii is computed, neglecting 
any radial transfer. The star-formation rate density is computed from the local gas density as
\begin{equation}
\Sigma_{SFR} = 0.00263 \, \Sigma_{GAS}^{1.48} \, V / R ,
\end{equation}
where $\Sigma_{SFR}$ is expressed in M$_{\odot}$ pc$^{-2}$ Gyr$^{-1}$, $\Sigma_{GAS}$ 
in M$_{\odot}$ pc$^{-2}$, $V$ in km/s and $R$ in kpc. 
This star-formation law was chosen because it reproduces observations in nearby galaxies
\citep{boissier03sfr}. The simulations of \citet{bush08} shows that spiral density waves can propagate
in an extended low surface density disk allowing star formation even at large disk radii.
The infall time-scale depends on mass and the surface density (dense regions are 
accreted early-on, massive galaxies are formed earlier than low mass galaxies). 
The models follow the chemical evolution of the disk, and calculate the resulting spectrum and 
colors.  They  also provide a
computation of the dust attenuation, based on simple assumptions to estimate the amount of dust from
the gas density and the metallicity. Since the assumptions concerning the dust role are very uncertain, we show our results
with and without dust attenuation. However,  at the very low density and 
metallicity considered here, the dust makes essentially no difference.

In this study of Malin 1, we need to extrapolate these models to much larger $\lambda$ values 
than it was done in previous studies. 
In \citet{boissier03lsb}, the spin parameter was investigated up to $\sim$ 0.2. Malin 1 is an extreme
case. We thus had to compute new models keeping all other assumptions identical, with 
very large values of the spin parameter (up to 0.7). 

\subsubsection{Best model for the extended disk of Malin 1}

We computed $\chi^2$ in a very similar manner as \citet{munoz11} to determine the best fit parameters and their uncertainties.
We performed this task within a radial range excluding the bumps clearly seen in the $FUV$ and $NUV$ profiles likely
linked to the spiral arms (in which the excess of light from recent star formation can outshine the longer-term 
star-formation history), and the bulge 
(which is especially pronounced at the redder wavelengths). 
We thus fit the models to match two radial range: 20-35 and 65-100 kpc. 
Different choices for the radial range have very little consequence on our final results as long as the two central points are excluded.
We performed the fit with three assumptions about the geometry as discussed in Section \ref{secgeompar} and 
using the photomery alone,  the gas profile alone, or both. The results are summarized in Table \ref{tabbestfit}.

\begin{figure*}
\centering
\includegraphics[angle=-90,width=19.cm,clip]{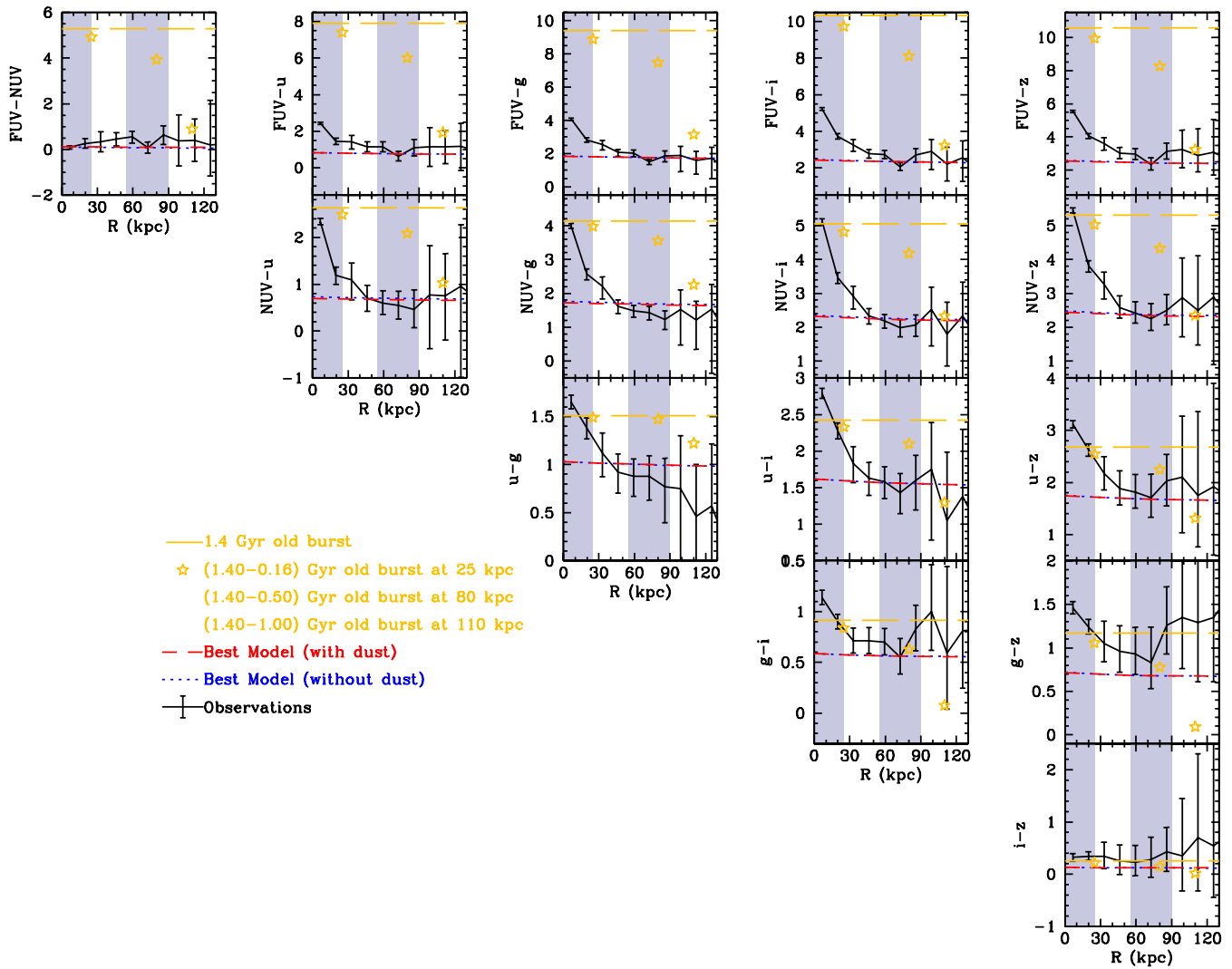}
\caption{All the color indexes profiles we have access to, and comparison to model predictions. 
The top-left side shows the color sensitive to young stellar population, while the bottom-right colors are sensitive to
older ones.
We show the best model discussed in Sect. \ref{sublargespin} (with our without dust) and the colors
of a burst of age 1.4 Gyr (horizontal line), and different ages for different radii (stars).
The shaded area show the regions excluded for the fit (bulge and bumps likely linked to the young populations of the spiral arm), but the color profile of the model is in full agreement with the observed one also in the remaining of the disk (within a few sigma).
}  
\label{figcolor}
\end{figure*}

\begin{table}
\caption{\label{tabbestfit}Best fitting models ($\chi^2$ based). }
\begin{tabular}{ l l l l}
\hline
\hline
b/a & used               & velocity (km/s) & spin  \\
PA        & constraints   & (km/s) & parameter \\
\hline
0.97                                     & photometry       &  490  (380 to 600)  & 0.61  (0.48  to  0.69)   \\ 
N/A                                       & gas          &  350  (340 to 360)  & 0.35  (0.32  to  0.39) \\
                                             & both         &  420  (360 to 540)  & 0.54  (0.45  to  0.67) \\
\hline
0.8 	                             & photometry           &  430  (360 to  580)  &  0.57 (0.47 to 0.69)	\\
40                                             & gas          &  330  (320 to  340)  &  0.39 (0.35 to 0.43)\\
                                             & both         &  380  (340 to  480)  &  0.51 (0.44 to 0.63)\\
\hline
0.8 	                                    & photometry   &   460 ( 370 to 600 )  & 0.58  ( 0.48 to 0.69) \\
var                                & gas                &   330 ( 320 to  340)  & 0.39  ( 0.39 to 0.43) \\
                                            & both              &   390 ( 350 to 390 )  & 0.51  ( 0.51 to 0.62) \\
\hline
\end{tabular}
\tablefoot{Fits were performed within a radial range excluding a bump related to 
the spiral structure and the bulge (see text). The best fit 
value of each parameter is given, along with the 68\% confidence interval computed as in Avni (1976).
The left column provides the geometry (b/a and PA). "var" means a variable PA has been adopted, following Lelli et al. (2010).}
\end{table}
Regardless of the assumptions we make, the profiles are best fit by a disk with a very large velocity and a very large spin parameter. 
We found that the lower limit of the parameters is better constrained (the $\chi^2$ increases quickly when moving towards lower values) than the upper limit. This is a strong constraint as it makes the extended disk of Malin 1 an 
extra-ordinary object, even if we adopt the 
lowest limit of the parameters interval allowed (i.e., a spin parameter larger than 0.3 and a velocity larger 
than 320 km/s). Since the gas profile does not
include the molecular component (even if it is likely to be low) and has a lower spatial
resolution, we prefer to adopt the parameters for the
fit of the photometric profile alone (the table clearly shows that similar values are obtained in any case).  
With the geometry of $b/a$=0.8 and a constant PA, we obtain $V$=430 km/s and $\lambda$=0.57, that we consider
our reference model in the following.

\subsubsection{Observed and modeled profiles}

The photometric profiles obtained for our best model are shown together with the observations in Fig. \ref{FigObsProfiles}
and the gas profile with the observed one in Fig. \ref{FigGas}.
\label{joel}
In this figure, we also show the stellar surface density of our best model. We compare it to the profile derived from the photometric profiles adopting the
method of \citet{roediger15}. The stellar mass to light ratio is estimated using the correlation found between the $i$-band mass to 
light ratio and $g$-$i$ color \citep{roediger15}
based on the FSPS stellar population model
\citep{conroy09}.
We assumed exponentially-declining star-formation histories, and a wide range of metallicities and reddenings.  
Despite the different method, and even the different assumptions used (e.g., for the IMF and star-formation histories), this method and our model provide
a consistent description of the stellar mass distribution in the giant disk of Malin 1.

Figure \ref{figcolor} shows the full set of color profiles obtained with our data.
The short wavelength colors (e.g., FUV-NUV) suggest recent or current star-formation activity,
as a relatively old burst alone could not explain the observed color. The colors in 
the redder bands (e.g., $i$-$z$) that are sensitive to the past history of star formation
suggest this activity has persisted for some time in the past. 
Our best model indeed reproduces the flat color gradient 
within the disk, except in the inner 20 kpc that 
are dominated by the bulge (that we do not attempt to model). 
If we follow \citet{barth07}, the inner 20 kpc resembles a normal 
galaxy, while the rest of the galaxy is the largest XUV star forming disk ever seen, that is surprisingly well fit
by a model entirely calibrated on usual disk galaxies such as the Milky Way.
We recognize that even if the best model is consistent with the colors profiles (for almost all the points within one sigma error-bars), some discrepancies are visible. For instance, the $FUV-z$ observed profile is a bit redder than the model one, and 
we can suspect a gradient in the inner 60 kpc. One possibility is that the inner part of the galaxy actually extends further out than the 25 kpc that we adopted as a limit. We have not attempted to subtract the extrapolation of an inner component that may affect the colors. The effect would be stronger for colors involving red bands (blue colors such as $FUV-NUV$, $FUV-u$ are quite flat, while as soon as we involve a red band starting from $g$, the central component becomes more prominent). This would affect both the gradient and the absolute value. Some systematic shifts may have other origins. For instance, the model is globally too red by a few tenths in $u-g$, and too blue in $g-z$ and $g-i$, what could be due to errors in the estimation of the sky level in the $g$ band. Since we are at very low surface brightness, the sky level uncertainty plays a big role. Despite our best effort, we may have a systematic offset due to an error in its determination. Our error-bars, however, include an estimate on this source of error (by measuring the sky in different part of the image around the galaxy and estimating its large-scale variation), and indeed the models are within these error-bars.
Finally, we used models with only 2 free parameters, while most of the galactic physics is fixed from previous works (e.g., the efficiency of star formation, the accretion history and its dependence on radius, the IMF). Some of the systematic differences could come from these assumptions (e.g., if the IMF varies slightly with density). By relaxing these assumptions, we could probably obtain better fits, but we would not learn much more about the galaxy.

Despite these small differences in colors, this simple model is consistent with the observed color profile, without fine-tuning outside of the two free parameters; and it predicts very different colors from other assumptions such as an old burst, or a star-formation ring expanding with time as discussed in Sect. \ref{secmapelli}.

\subsubsection{Predictions concerning the History of the Malin 1 giant disk}

\begin{figure}
\centering
\includegraphics[width=9.cm,clip]{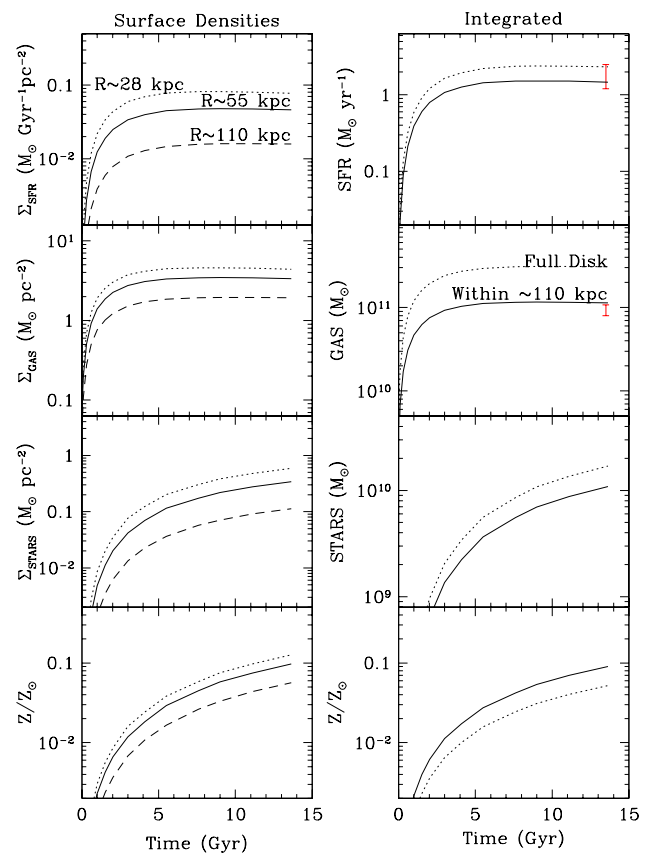}
\caption{History of the extended disk in Malin 1 according to our best-fit model. Left: evolution at 
specific radii (28, 55, and 110 kpc). Right: evolution of integrated quantities 
(over the full disk modeled, or within about 110 kpc, corresponding to the size of the 
visible disk). The integrated values are compared to
the values of \citet{boissier08} for the star-formation rate and \citet{lelli10} for the gas
(shown by the  errorbar in the topmost two right panels).}
\label{FigHistory}
\end{figure}

Under the assumption that this model is correct, we can make some predictions on 
the history of the galaxy, and on its past
and present metallicity (Fig. \ref{FigHistory}).

A long, protracted star-formation history is necessary to account for the galaxy's large 
mass: the low stellar density and large spatial extent, combined with the adopted star-formation law, 
implies a low star-formation efficiency and a relatively modest star-formation rate that gradually builds 
up the disk of the galaxy over an extended period of time.
This long history is also consistent with the kinetic information. Indeed, the HI kinematics is regular and symmetric, while 
the orbital time scale is extremely long ($\sim$ 3 Gyr). Then, the outer disk must have been in place 
and undisturbed for many Gyrs \citep{lelli10}.
The resulting stellar mass to light ratio in the extended disk is low, as found in blue galaxies
\citep[see e.g., in][]{kauffmann03,gallazzi09} with values of $log(M_{\star}/L)$ around $-0.45$ in the $u$ band, $-0.3$ in the $g$ band
and $-0.2$ in the $i$ and $z$ bands. The baryonic mass (including gas) to light ratio instead rises
from $log(M/L)\sim$ 0.4 to 0.7 between the center and R $\sim$ 100 kpc in the $u$ band (0.55 to 0.9 in $g$ band,  0.65 to 1 in $i$ and $z$ bands).
 This scenario also implies a low gas metallicity 
(slightly lower than 1/10 solar), which is consistent with the metallicities usually measured in LSBG, that are lower 
than in high surface brightness galaxies \citep[e.g.,][]{mcgaugh94,deblok98,liang10}.

\subsubsection{Discussion on the velocity and spin}

Our reference model for the geometry of $b/a$=0.8 and a constant PA has for parameters
$V$=430 km/s and $\lambda$=0.57 (similar values are found under the other geometries that we have tested).
This velocity is much higher than the circular velocity found by \citet{lelli10} ($V\approx$220 km/s from the HI kinematics) although 
they adopted a similar inclination. 
This inclination is, however, quite uncertain since it is based purely on a manual fit 
of an ellipse to the $R$ band
image of \citet{moore06}. 
Even a very small error on the inclination would translate in a large error in the velocity since the galaxy is almost face-on.
As a result, the velocity of \citet{lelli10} is itself very uncertain. 
If the disk is more inclined than assumed by \citet{lelli10}, then the velocity should be corrected for the different 
inclination. They adopted an inclination of 38 degrees. If the inclination is in fact 14 degrees (corresponding 
to the axis ratio of 0.97 that was measured in the $u$ band), the velocity of \citet{lelli10} corrected for the inclination becomes 564 km/s. Our best model is then instead $V$=490 km/s  and $\lambda$=0.61 (likely in the range 0.35-0.7). 
A giant rotating disk with velocity in the range 300 to 500 km/s and spin in the range 0.35-0.7 is thus consistent with all the observations. 
This velocity is consistent with the observations of \citet{lelli10} only if the inclination is 18 degrees (instead of the 38 degrees they assumed). 
Galaxies with velocities around 300-400 kms/s are rare at best. Some may exist if we take
at face values the velocity functions of \citet{gonzalez00} who combined luminosity functions and Tully-Fisher relations, and no galaxies are known with velocity larger than 300 
\citep[][from resolved rotation curves of 175 galaxies]{lelli16} or 400 km/s \citep[from the HI line-width of the 4315 detections in the HIPASS catalog by][]{zwann10}.

The spin parameter is also extremely large. For comparison, the properties of the Milky Way are well 
reproduced by a model with
$V$ $\sim$ 220 km/s and $\lambda \sim$ 0.03, and 
\citet{munoz11} showed that the photometric profiles of 42 spiral galaxies of the SINGS sample
were well  reproduced using the same models with velocities in the range 50 - 300 km/s, and $\lambda$ in the range 0.02 - 0.08 
except for two galaxies with $\lambda$ $\sim$ 0.14. 
For a larger number of galaxies but with a different method,  
\citet{hernandez06} also found a very similar spin parameter distribution.

\begin{figure}
\centering
\includegraphics[width=9.cm,clip]{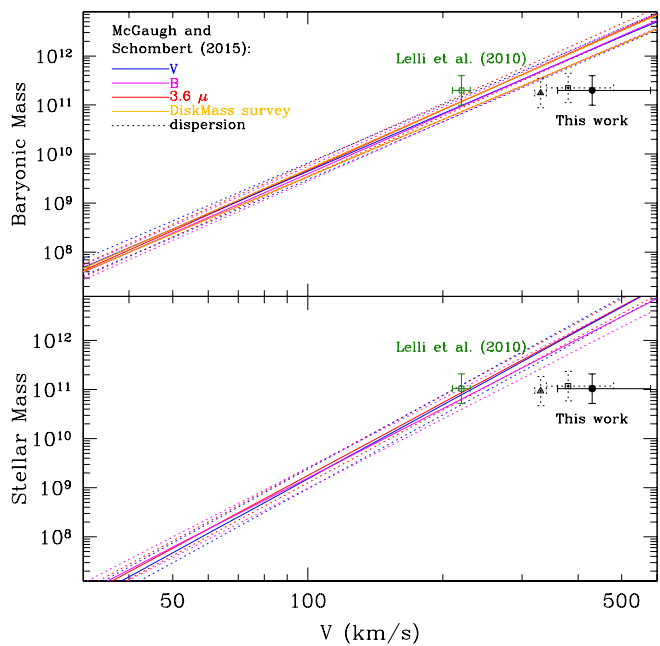}
\caption{Stellar and Baryonic  Tully-Fisher relationships taken from \citet{mcgaugh15}. The 
different lines indicate different samples and wavelengths used in their work. The position of Malin 1 for the 
$\sim$ 220 km/s reported in \citet{lelli10} is indicated, as well as the ``velocity'' derived from 
our model fit (filled circle, open triangle, and open square for respectively the fit based on the photomeric profiles,
the gas profile, and both).}
\label{FigTF}
\end{figure}

The fact that Malin 1 requires extreme values to be fit may not be that surprising, since we already know
that it is the most extreme galaxy for instance in central surface brightness and scale-length \citep[see Fig. 7 of][]{hagen16}. 
We should thus keep in mind the possibility that we need an extended range of physical parameters with respect 
to other galaxies. 
On the other hand, these extremely large values suggest 
that this model may not be physical. Indeed the
model implements scaling relations (between velocity, spin parameters, mass and scale-lengths) that may break 
down for the  extended disk.
Figure \ref{FigTF} indicates that it might be the case since the velocity that we obtained with our fit
is too large for the galaxy as a whole to be on the Tully Fisher relationship (either the stellar one,
or the baryonic one, probably more appropriate for this gas rich galaxy). 
The stellar mass was estimated by summing the stellar mass we measure in the
extended disk, and the stellar mass in the inner disk as derived from the magnitude published in 
\citet{barth07} combined with the stellar mass to light ratio in the R band of 3.3 of \citet{lelli10}.
For the baryonic relation, the mass of gas that needs to be added is obtained by multiplying the HI mass
of \citet{lelli10} by a 1.4 factor to take into account other species. The largest velocities suggested by our
fit is indeed  too large with respect to the baryonic Tully Fisher relationship, unless Malin 1 is an outlier.
Intermediate values around 300 km/s would still agree well with the relationship.

If this is a good indication that the usual scaling laws break down in the extended disk of Malin 1, then
the   velocity and the angular momentum corresponding to the best fit loose their physical sense. 
The relative success of our model still may indicate that the extended disk of Malin 1 behaves like it 
is expected for a normal disk of low density : 
i) the model produces a correct star-formation law for the observed densities and radius, 
ii) it produces a correct star-formation history with right colors and stellar light density, 
iii) the extended disk follow an about exponential distribution.


We also made some ad-hoc models to understand
the role of diverse assumptions in our results. We fixed the total surface density radial profile from the 
stellar and gas profiles (as seen in Fig. 6), and the rotation curve to be the one derived
by \citet{lelli10} instead of using the scaling laws. Using the same star-formation law, and dependence of the accretion history on the 
velocity and surface density as in previous models, we found that the model does not provide
a good fit to the data. While the gas profile is very similar to the observed one, we find two discrepancies: 
1) the model under-estimates the surface brightness at radii larger than about 100 kpc (by more than 1 mag at 120 kpc),
2) the model predicts systematically bluer colors than the observed ones.
Point 2 is related to the accretion history that we have adopted, 
corresponding to the low density and a  velocity of about 200 km/s: 
most
of the accretion occurs very late in the history of the galaxy. The light is then dominated by very young blue stars.
We performed new computations with various accretion histories (taken from the usual models for different velocities),  and various star 
formation efficiencies (adding a multiplying factor in equation 1). 
The colors are systematically improved when the accretion occurs earlier than in the first model.  
Modifying the efficiency has 
little effect on the colors. The best model is found by adopting a reduced efficiency (using a factor 0.5)  
but a very low efficiency (a factor lower than 0.3) and a very large one (larger than 1.5) are excluded.  The star-formation law of equation 1 being deduced from data in which the scatter is close to a factor 2, this deviation in efficiency does not significantly distinguish Malin 1 from 
regular galaxies.
Concerning the point 1, surface brightness profiles in better agreement with the data 
could be obtained by rising the star-formation efficiency with radius, but  it is not clear on which physical basis this
 should occur. Another possibility is that our census of stars and gas is incomplete (e.g., wrong $M/L$ ratios, hidden gas at large radii). We tested this by correcting the total surface density by a factor corresponding to the difference in surface brightness 
between our best ad-hoc model and the observations, and running the grid of models again. 
The surface brightness profiles were improved, but the gas profile was then poorly fit. 
The color profiles were almost identical as in the previous set of models.
Finally, all of the models constructed by imposing the observed surface density profile and rotation curve
have worse $\chi^2$ than the one presented before. This is probably because the scaling
relations with $V$ and $\lambda$ allowed us to explore a larger variety of surface densities, and thus of star-formation histories.
The main conclusion of these tests is that an early accretion of the extended disk of Malin 1 is always favored. 
This is consistent with the other indications  of a relatively long history of the giant disk as discussed before.


\subsection{Ring galaxies and LSBGs}
\label{secmapelli}

\citet{mapelli08} proposed that ring galaxies could be the ancestor of LSBGs. 
Ring galaxies exhibit a peak moving progressively outwards with time
in their surface brightness profiles \citep{vorobyov03}.
The simulations of 
\citet{mapelli08} suggest that galaxy collisions indeed create rings that are visible up to about 0.5 Gyr
after the interaction. They propose a model for Malin 1 where the collision occurred 1.4 Gyr ago
for the galaxy to have the time to expand to such a size and the ring to dissolve into a large disk.
Their work was based on a
comparison with the \citet{moore06} data. In our new surface photometry  
profiles, we do not see the typical features in the stellar density profile that they predict, 
such as a break around 50 kpc.
In our images (Fig. \ref{FigObsNGVSnative} and appendix A), we do not see  
any sign of stellar "spokes" like those seen in the Fig. 4 of \citet{mapelli08} for their detailed model of Malin 1, or for other LSBGs 
\citep[such features are not visible in the $g-$ and $r-$band images of][either, even if the authors applied special techniques to enhance the contrast of structures]{galaz15}.
Some significant number of stars should also have been created during the ring phase. Even if they do
not dominate the stellar population, they should have left some traces in the images, in the profiles,
or in the SEDs of some regions including stars created at that time.

In Fig. \ref{figcolor}, we placed a horizontal line corresponding to the color of a 1.4 Gyr old burst,
adopting the \citet{boissier08} simple model for the colors of aging stellar populations.
In their scenario, \citet{mapelli08} suggest the ring is located at about 25, 80, and 110 kpc, 
respectively, 0.16, 0.5 and 1 Gyr after
the interaction. We thus indicate in the same figure the color expected for the remnant of the stars formed at that times, assuming they 
still dominate the stellar light at the corresponding radius. Clearly, the color profiles in Malin 1 are not in good agreement with this scenario.
If stars were formed during the collision and the ring phase, they have left no obvious traces in the images or in the profiles.
The ages of individual regions that we defined and fit with a single stellar population  
(or with more complex star-formation histories) in Sect. \ref{secSEDregions}
are spread over a large range of values that do not agree particularly well with the ages expected based on the model.

We thus conclude that the color 
gradient in Malin 1 is not consistent with the progenitor being a ring galaxy. 
Instead,  a long-term low efficiency 
formation of stars  in a large disk,  as presented in the previous section, is more likely.

\subsection{Intermittent star-formation histories}

The fits of selected regions along the spiral arms suggest different ages, spread over several 100 Myr, with signs of older stellar 
population (several regions e.g., 6,14,16) have red SEDs that are not well fit by a single recent population). We found no age 
gradient to which the spread of ages could be correlated.

This may indicate that, while the general model discussed in Section \ref{sublargespin} 
represents the long-term regular history of the galaxy, the recent evolution is 
characterized by quiescent and active star-formation periods (so that we find regions in these different phases), 
as was suggested from the $FUV-NUV$ integrated color of 
a sample of LSBG galaxies in \citet{boissier08}. 
Variations in the star-formation histories could also explain the existence of optically 
red LSBGs \citep{boissier03lsb}.
The fact that some regions are blue in $FUV-NUV$ but red in other colors could also be explained if the blue color is due to a recent event of star formation in the last 100 Myr (contributing mostly to $FUV$), while the 
star formation on NUV emission time-scale (500 Myr) 
was lower, and if a significant amount of stars was formed 
at earlier epochs.

An optical spectrum with an estimate of the \halpha{} emission is available only for the center 
of the galaxy \citep{impey89}. It would be extremely interesting to obtain new data on line emission in the
galaxy to test the micro-variations of the star-formation history on time scales shorter than 100 Myr. 
New instruments,
such as MUSE, with excellent efficiency and sufficiently large field of view should allow us to achieve this goal soon.

Our results then globally indicate that the properties of the extended disk of Malin 1 are consistent with a long history of 
low level star formation, with increased activity at some epochs in different regions, separated by phases 
that can last several 100 Myr. This is quite consistent with the analysis of the stellar populations in a sample of 
LSBG galaxies performed by \citet{schombert14}, and with the results of \citet{zhong10} for blue LSBGs.
From HST data, \citet{vallenari05} also found that UGC5889, another LSBG  has experienced modest bursts and
very quiescent periods.
Such a scenario for the star-formation histories is also suggested by the
N-body simulations of \citet{gerritsen99}.

%


%

\section{Conclusion}
\label{secconclu}
We present a panchromatic view of Malin~1 based on deep UV-optical images from
the NGVS and GUViCS surveys. The six photometric bands ($FUV$, $NUV$, $u$, $g$, $i$, $z$) 
used in our analysis are crucial for determining the  stellar population of the galaxy and have 
allowed us to reach several notable conclusions concerning the properties of this prototypical
low surface brightness giant galaxy.

\begin{itemize}
\item The extended disk of Malin 1 is consistent with a long standing disk that has  been forming stars with 
a low star-formation efficiency. 

\item This low surface brightness disk is well fit by a simple model of evolution
of disk galaxies. In this model the only significant difference with the Milky Way is 
the angular momentum, that is exceptionally large in the case of Malin 1:
its spin parameter about 20 times larger than that of the Milky Way. 
The model however relies on scaling relationships that may break down
in the extended disk of Malin1. The large velocities and spin parameter obtained
may in that case loose their physical meaning.
The predicted evolution could, however, still apply to the low density disk.
This model allows us to make predictions
on the stellar population of the galaxy, and the metallicity of the disk. Such predictions are
testable by future observations. To maximize the consistency with the observations,
we suggest the inclination of the galaxy could be around 18 degrees 
(or lower than 40 degrees
often adopted in the literature) although this value is not strongly constrained.

\item The stellar disk of Malin 1 extends at least out to 130 kpc in radius, 
confirming Malin 1 as 
a candidate for the title of largest known galaxy in the universe (and leaving open the question 
about how such large structures can survive during the hierarchical evolution of structures in a 
cold dark matter dominated universe).

\item The morphological structure and colors of the extended disk in Malin 1 show no evidence 
of Malin 1 having a ring galaxy as ancestor, as previously suggested.

\item Intermittent star-formation histories are favored by the colors of  individual regions, as was
suggested in some previous work on LSBGs.

\end{itemize}

Some of the conclusions reached in this paper could  apply to other giant LSBGs, Malin 1 being the most
spectacular case of this class of galaxy \citep[although our current census of Malin 1 analogs in the local 
volume may be incomplete, ][]{impey89}. Likewise, some findings may be relevant to the numerous XUV disks found
around normal disk galaxies \citep{thilker07}.

\begin{acknowledgements}
We thank the referee for pointing to very pertinent elements, allowing us to improve our analysis and discussion.
This work is supported in part by the Canadian Advanced Network for Astronomical Research (CANFAR) which has been made possible by funding from CANARIE under the Network-Enabled Platforms program. This research used the facilities of the Canadian Astronomy Data Centre
operated by the National Research Council of Canada with the support of the Canadian Space Agency.
JK acknowledges the support from NASA through grant NNX14AF74G.
This work is based in part on data products produced by GAZPAR located at the Laboratoire d'Astrophysique de Marseille. 
We especially thank Denis Burgarella and Olivier Ilbert for their feedback and help in this process.
This research has made use of the NASA/IPAC Extragalactic Database (NED) which is operated by the Jet Propulsion Laboratory, California Institute of Technology, under contract with the National Aeronautics and Space Administration. 
This research made use of Montage. It is funded by the National Science Foundation under Grant Number ACI-1440620, and was previously funded by the National Aeronautics and Space Administration's Earth Science Technology Office, Computation Technologies Project, under Cooperative Agreement Number NCC5-626 between NASA and the California Institute of Technology.
\end{acknowledgements}

   \bibliographystyle{aa} 
   \bibliography{/home/boissier/Ddata1link/Dutil/Dlatex/Dbibs/ALLREFS} 
%

\newpage

\begin{appendix} 

\section{Panchromatic images of Malin 1 at their native resolution.}

\begin{figure*}
\centering
\includegraphics[width=18.cm,clip]{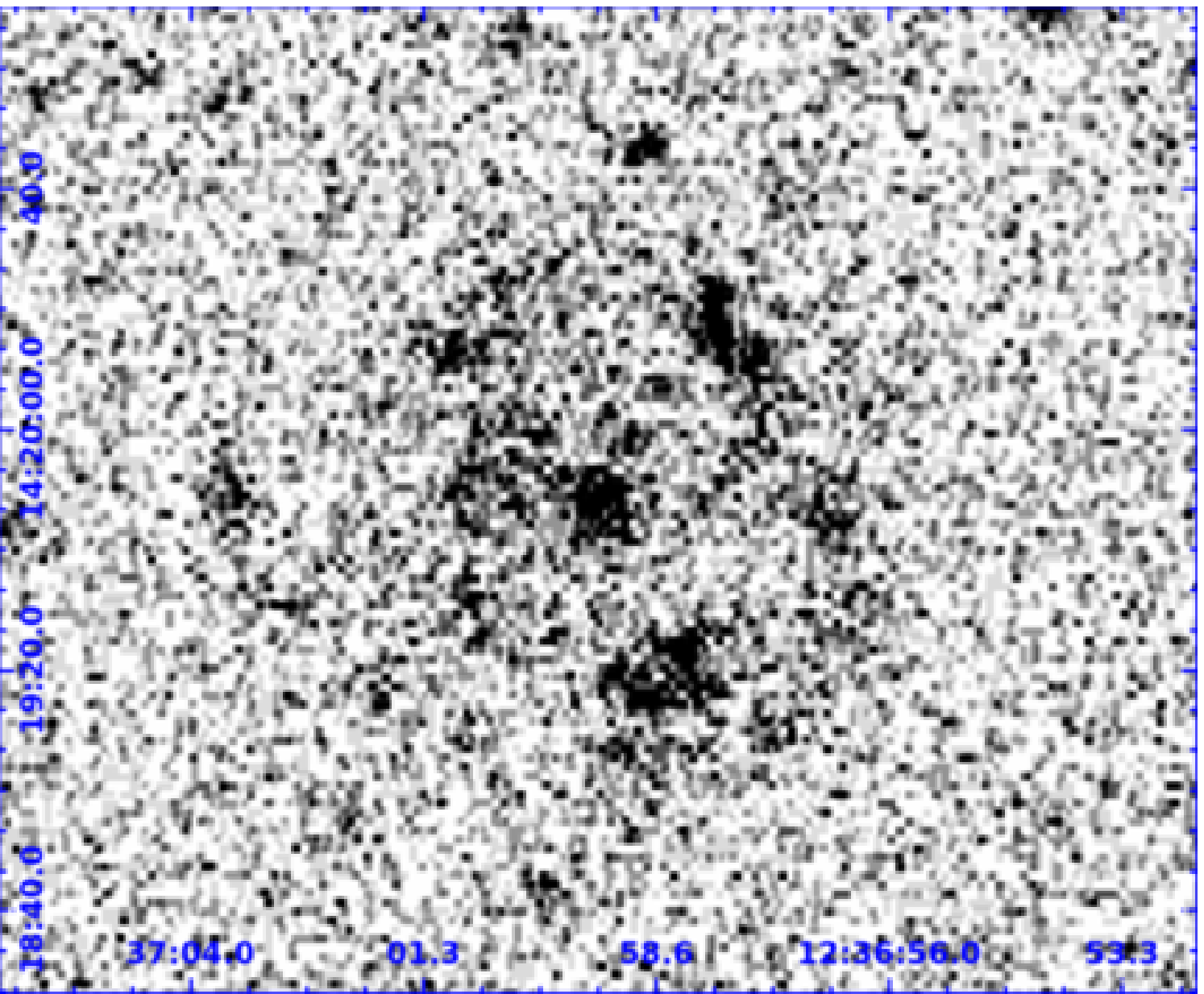}
\caption{$FUV$-band image.}
\label{figbandfuv}
\end{figure*}

\begin{figure*}
\centering
\includegraphics[width=18.cm,clip]{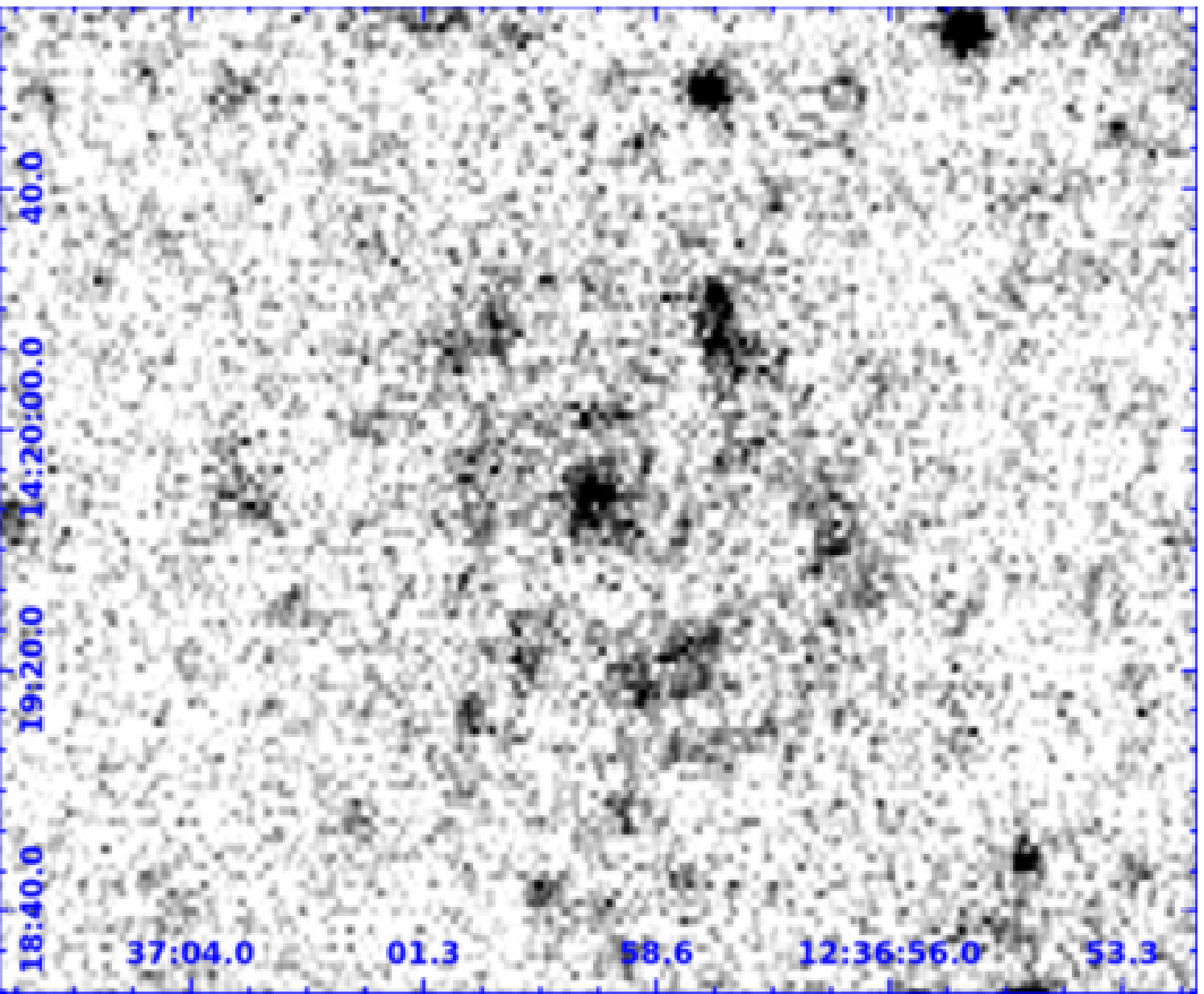}
\caption{$NUV$-band image.}
\label{figbandnuv}
\end{figure*}

\begin{figure*}
\centering
\includegraphics[width=18.cm,clip]{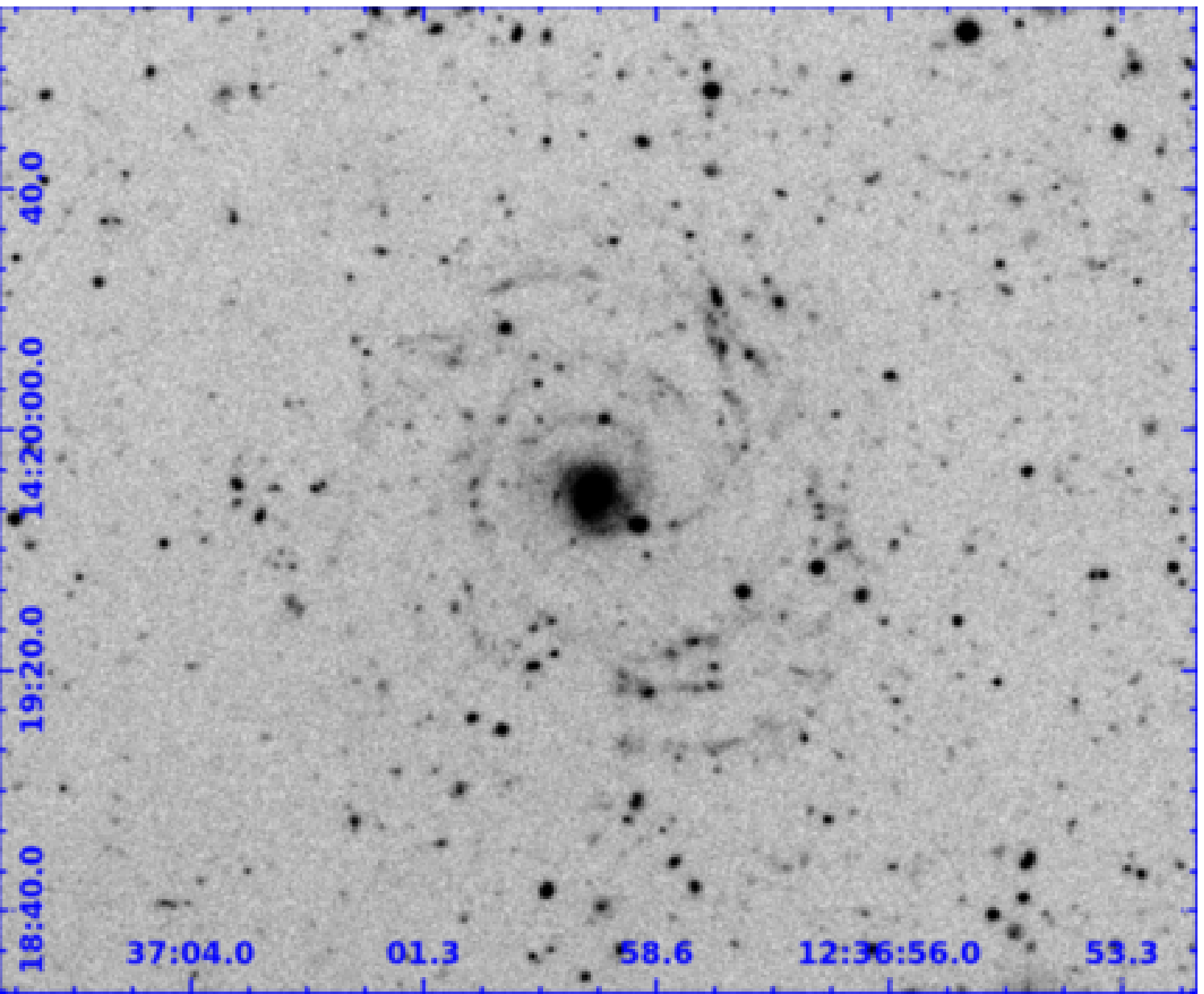}
\caption{$u$-band image.}
\label{figbandu}
\end{figure*}

\begin{figure*}
\centering
\includegraphics[width=18.cm,clip]{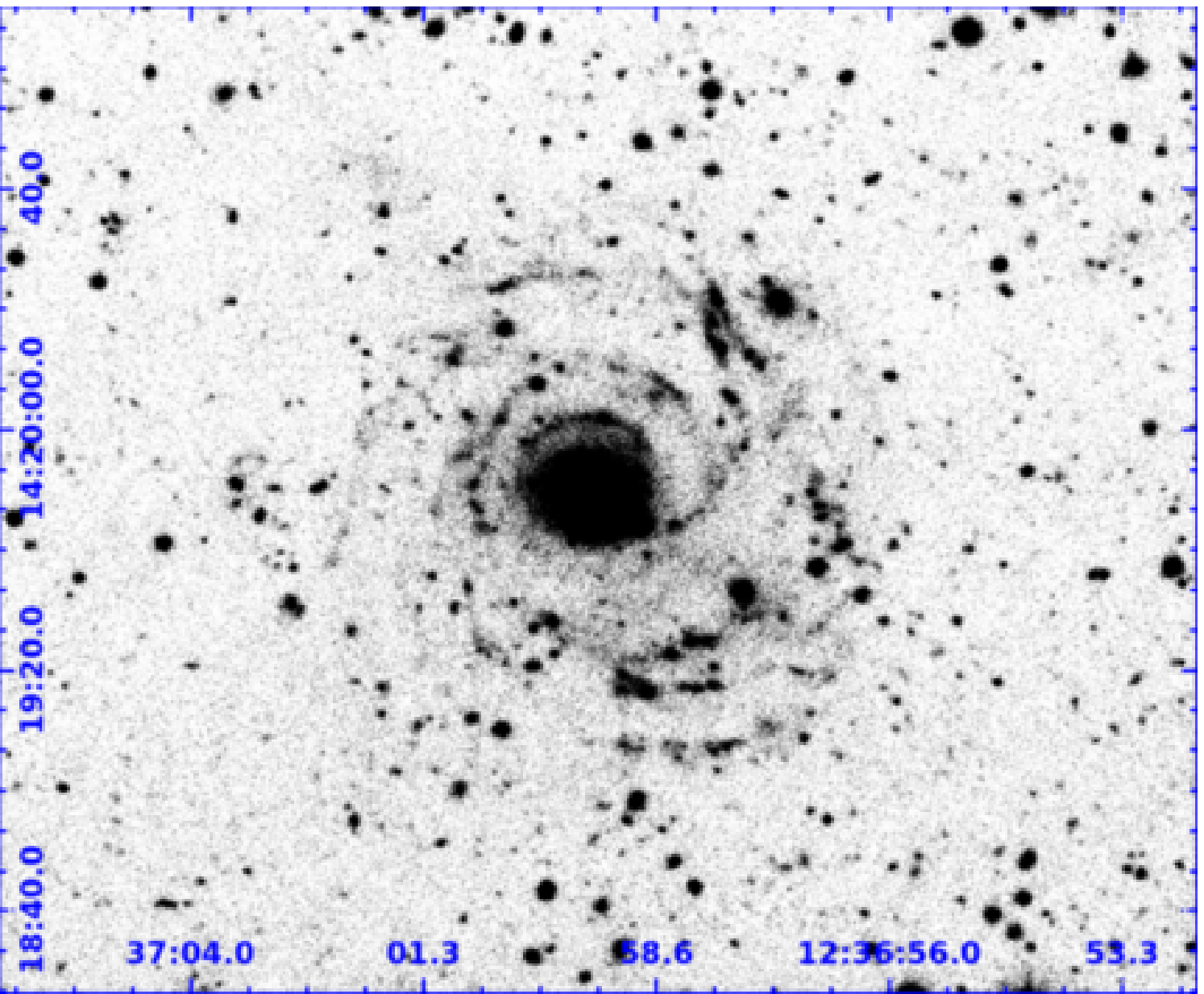}
\caption{$g$-band image.}
\label{figbandg}
\end{figure*}

\begin{figure*}
\centering
\includegraphics[width=18.cm,clip]{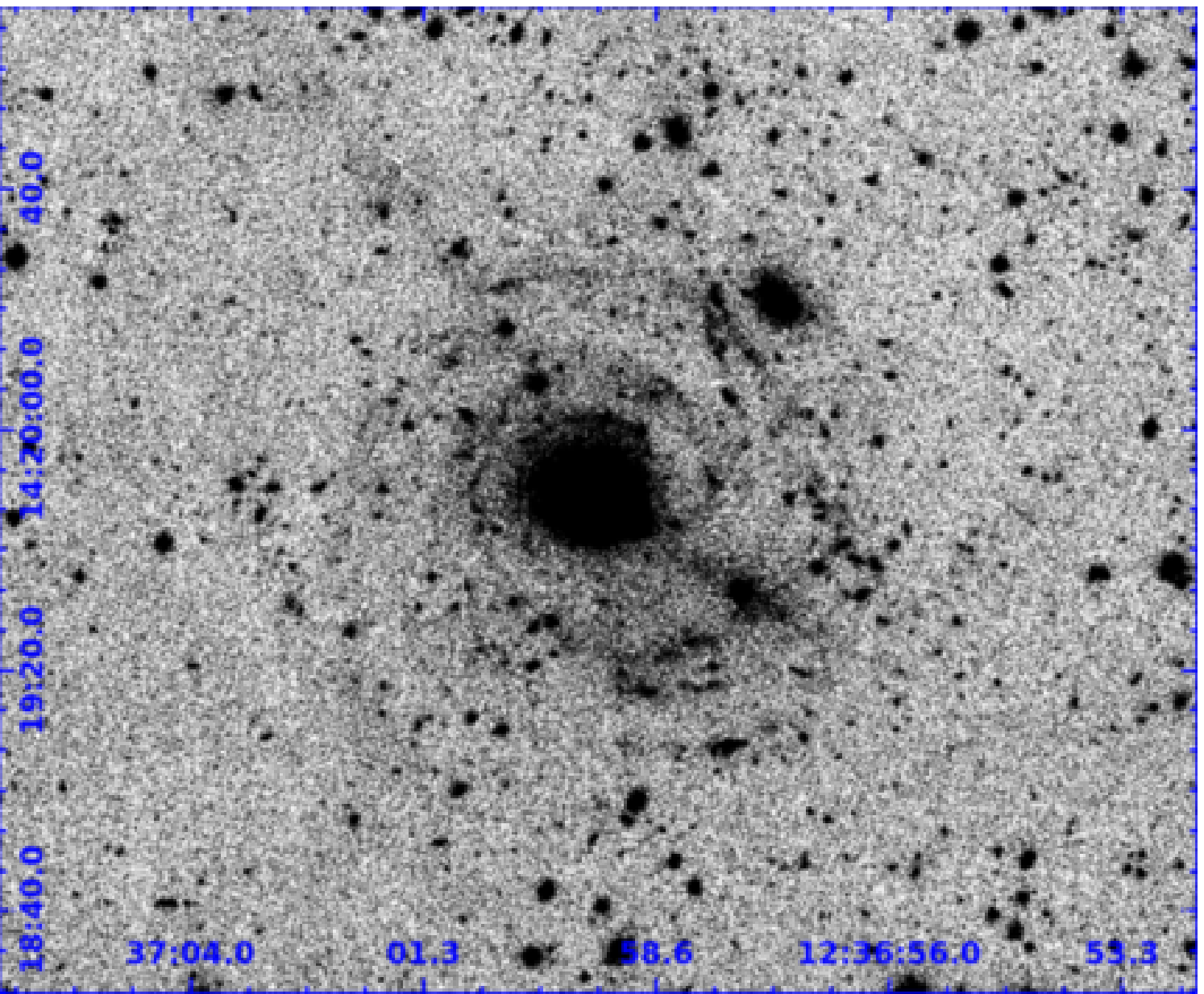}
\caption{$i$-band image.}
\label{figbandi}
\end{figure*}

\begin{figure*}
\centering
\includegraphics[width=18.cm,clip]{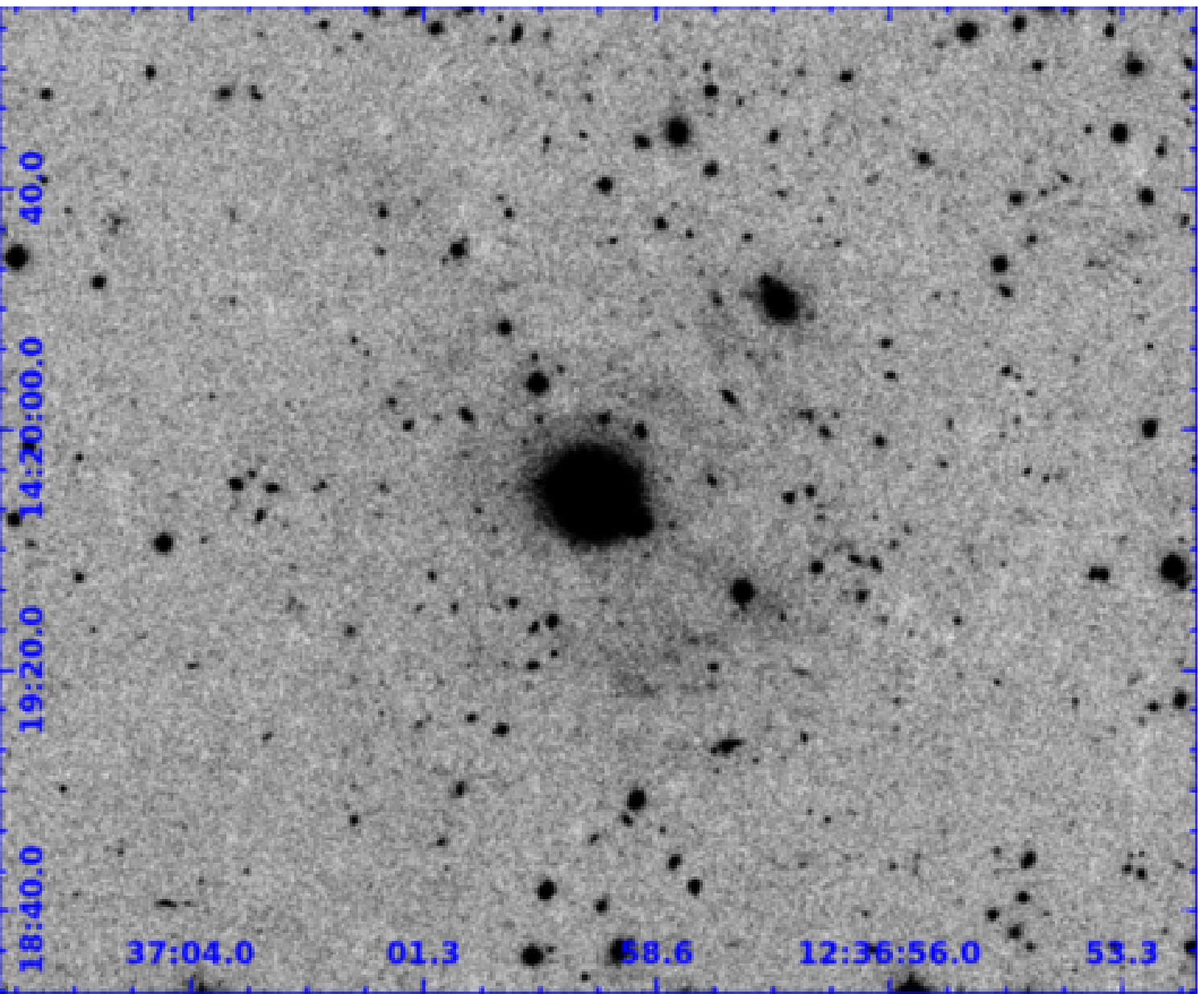}
\caption{$z$-band image.}
\label{figbandz}
\end{figure*}

Figures  \ref{figbandfuv} to \ref{figbandz} show the six images at their native 
resolution. 
\label{secallimages} 

\section{GAZPAR modeling of the 16 regions spectral energy distributions.}
\label{secgazparfit}
\begin{figure*}
\centering
\includegraphics[angle=-90,width=18.cm,clip]{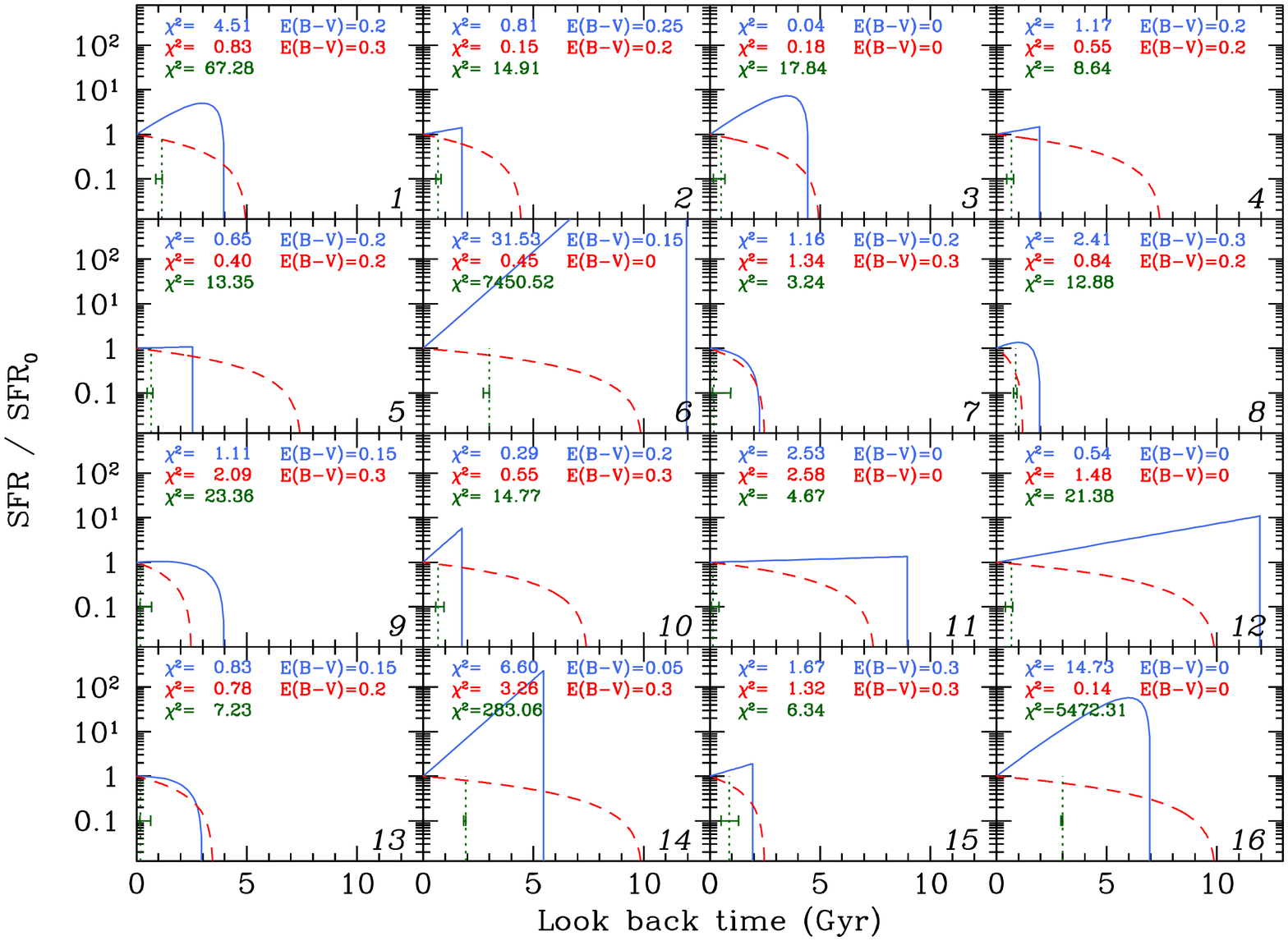}
\caption{Star-formation history of individual regions for different models. The vertical dotted line
indicates the age of a single population. The error-bar indicates the uncertainty on the age. 
The red dashed curve indicates the best star-formation history found by CIGALE. The blue solid curve indicates
the best star-formation history found by Le Phare. In each panel, the non-reduced $\chi^2$ is given (from top to bottom: Le Phare, CIGALE, 
single generation). In the first two cases, the fit allows for the presence of dust, and the reddening of the best model is also given.}
\label{FigSFHregions}
\end{figure*}

GAZPAR allows
us to fit the SEDs of the regions selected in Malin 1 using two tools widely used in the community to compute for instance photometric redshifts or physical parameters, proposing different level of choices for various parameters and assumptions.

We first used GAZPAR to fit our data to Le Phare-Physical parameters. 
General presentations of this code can be found in \citet{arnouts99,ilbert06}.
The SEDs are fit with standard templates 
corresponding to Bruzual \& Charlot (BC03) single stellar populations combined with a set of diverse star-formation histories 
(exponentially declining or 
delayed), a Chabrier IMF, $E(B-V)$ from 0 to 0.3 in 0.05 steps. GAZPAR uses the Chabrier IMF, 
while the models discussed in the  main part of this work use the \citet{kroupa01} IMF. However both IMFs are very similar for our purpose.

We then submitted the same catalog to the CIGALE code through the same interface. 
General information on CIGALE can be found in \citet{burgarella05} and \citet{noll09}.
We also use BC03 stellar population
and Chabrier IMF. We choose delayed star-formation histories for which ages and time-scales have to be defined by the user. 
We allowed a large number of values for the timescale (in Myr): 10, 20, 50, 100 to 1000 in steps of 100, 1000 to 2000
 in steps of 200, 2000 to 5000 in steps of 500, creating a variety of star-formation histories.
We allowed the same values for possible ages, with the addition of even older ages (7500, 10000) to allow for stellar 
populations spread over a large range of ages. 
CIGALE allows the choice of a low  metallicity (0.004), which we adopted since LSBGs are likely to be un-evolved
\citep{mcgaugh94,deblok98,burkholder01,liang10}. No metallicity data are however available for Malin 1 yet.

The results of both methods are show in Fig. \ref{FigSFHregions} where we also indicated the 
age obtained in Sect. \ref{secburst} for a single star forming event.

We remark that the two GAZPAR codes allowing to fit for an attenuation are in a surprisingly good 
agreement for this parameter in most regions. 
This does not necessarily imply that the result is telling us about the dust (both codes are based on similar principles, 
and they could use attenuation as a free parameter to compensate for other short-comings of our approach). 
However it suggests the regions with largest attenuation ($E(B-V) \sim$ 0.2 - 0.3) found by both codes 
(regions 1, 7, 8, 10, 15)
could be interesting for future observations of the far infrared dust emission, or of molecular gas tracers.

In most regions, the two GAZPAR codes provide a consistent star-formation history. It indicates for several 
of them (1, 8, 9, 13, 15) short ages (and/or an increasing star-formation history with time). In all of the regions, the age
derived in the simpler model of a fading burst is shorter than the age derived by the more complex star-formation history.
All of this indicates that the luminosity of these regions is dominated by recent star formation. The two GAZPAR
codes also agree in several regions to suggest an extended star-formation history 
(6, 11, 12, 14, 16) showing that a relatively 
old stellar population exists. This is true for regions 6, 14, 16 
that were chosen as blue  from their $FUV$ emission. 
The $\chi^2$ for 
those regions are not good in any scenarios. The blue $FUV-NUV$ color could be due to a recent and weak star forming event, on top 
of an old underlying stellar population responsible for the red optical colors that are indeed visible in Fig. \ref{FigSedRegions}.
In some regions, the two codes provide different histories --- a 
declining star-formation history, but of young age for Le Phare,
and a rising star-formation history for CIGALE. This is likely due to 
regions where the light is dominated by a relatively young
stellar population (less than few Gyr) as both scenarios indicate a high 
number of young stars.

\end{appendix}
\end{document}